\pdfoutput=1
\documentclass[10pt,twocolumn,letterpaper]{article}

\usepackage[letterpaper,textwidth=6.75in,textheight=9in,columnsep=0.25in,hcentering,top=1in,
            headheight=0pt,headsep=0pt,footskip=0.4in]{geometry}
\usepackage{times}
\usepackage{microtype}
\usepackage{graphicx}
\usepackage{booktabs}
\usepackage{amsmath}
\usepackage{xcolor}
\usepackage{hyperref}
\usepackage{natbib}
\usepackage{xurl}
\usepackage{enumitem}

\usepackage{xspace}
\usepackage{array}

\newcommand{\sys}{\textsc{Callo}\xspace}
\newcommand{\sysstream}{\textsc{Callo}-Stream\xspace}
\newcommand{\base}{vLLM\xspace}

\newcommand{\invkv}{invalidated KV\xspace}
\newcommand{\Invkv}{Invalidated KV\xspace}
\newcommand{\provkv}{provisional KV\xspace}

\newcommand{\mainmodel}{Llama-3.3-70B\xspace}
\newcommand{\mainwl}{SWE-agent L5\xspace}

\newcommand{\fixedtie}{fixed-tie greedy\xspace}
\newcommand{\anymax}{any-max greedy\xspace}
\newcommand{\Anymax}{Any-max greedy\xspace}

\definecolor{linkblue}{rgb}{0,0.08,0.45}
\hypersetup{
  pdftitle={Callo: Serving Cache-Miss Requests with Speculative Decoding on Invalidated KV under Prefill-Decode Disaggregation},
  pdfauthor={Jianjun Ou, Matthew Lentz},
  pdfborder=0 0 0,
  pdfpagemode=UseNone,
  colorlinks=true,
  linkcolor=linkblue,
  citecolor=linkblue,
  filecolor=linkblue,
  urlcolor=linkblue}

\setcitestyle{authoryear,round,citesep={;},aysep={,},yysep={;}}
\renewcommand{\cite}[1]{\citep{#1}}
\makeatletter
\renewcommand\small{\@setfontsize\small\@ixpt{10}}
\renewcommand\footnotesize{\@setfontsize\footnotesize\@ixpt{10}}
\renewcommand\scriptsize{\@setfontsize\scriptsize\@viipt{8}}
\renewcommand\tiny{\@setfontsize\tiny\@vipt{7}}
\renewcommand\large{\@setfontsize\large\@xiipt{14}}
\renewcommand\Large{\@setfontsize\Large\@xivpt{16}}
\renewcommand\LARGE{\@setfontsize\LARGE\@xviipt{20}}

\renewcommand\section{\@startsection{section}{1}{\z@}{-0.12in}{0.02in}{\large\bfseries\raggedright}}
\renewcommand\subsection{\@startsection{subsection}{2}{\z@}{-0.10in}{0.01in}{\normalsize\bfseries\raggedright}}

\def\@listi{\leftmargin\leftmargini}
\def\@listii{\leftmargin\leftmarginii
  \labelwidth\leftmarginii \advance\labelwidth-\labelsep
  \topsep 2pt plus 1pt minus 0.5pt
  \parsep 1pt plus 0.5pt minus 0.5pt
  \itemsep \parsep}

\renewcommand\footnoterule{\kern-3pt \hrule width 0.8in \kern 2.6pt}

\newsavebox\captionlabelbox
\long\def\@makecaption#1#2{%
  \vskip 10pt
  \baselineskip 11pt
  \setbox\@tempboxa\hbox{#1. #2}%
  \ifdim \wd\@tempboxa >\hsize
    \sbox{\captionlabelbox}{\small\slshape #1.~}%
    \usebox\captionlabelbox {\footnotesize #2}%
  \else
    \centerline{{\small\slshape #1.} {\small #2}}%
  \fi}

\renewenvironment{abstract}
  {\centerline{\large\bfseries Abstract}\vspace{-0.12in}\begin{quote}}
  {\par\end{quote}\vskip 0.12in}
\makeatother

\begin{document}

\twocolumn[{%
\begin{center}
{\Large\bfseries\baselineskip 18pt Callo: Serving Cache-Miss Requests with Speculative Decoding\\on Invalidated KV under Prefill-Decode Disaggregation\par}
\vskip 0.25in
\begin{tabular}[t]{c@{\hspace{0.9in}}c}
\textbf{Jianjun Ou} & \textbf{Matthew Lentz} \\
Duke University & Duke University \\
\texttt{jianjun.ou@duke.edu} & \texttt{mlentz@cs.duke.edu}
\end{tabular}
\end{center}
\vskip 0.2in

\begin{abstract}

Agent sessions may involve frequent context edits, which effectively invalidate cached KV state and force requests to wait for re-prefill before decoding.
We present \sys, a disaggregated serving system that leverages invalidated KV for speculative drafts to reduce this latency, while verifying drafts against the exact KV after prefill before committing tokens.
\sys makes effective use of decode resources by scheduling speculations during periods of underutilization and preempting speculations when drafts are likely to not be accepted.
We evaluated \sys on a two-GPU testbed using SWE-agent sessions across three model families, observing all-request speedups of 1.05--1.19$\times$ over vLLM at moderate load.
We demonstrate that these gains are also realized at scale through cluster-level simulations driven by real-world traces.

\end{abstract}

}]

\section{Introduction}

Agents frequently work in long sessions~\cite{tracelab} that send many requests to the model, where each request includes prior session context as input.
For example, coding agents read files, edit code, and run tests within a single session.
Across 4,764 real-world Claude Code sessions~\cite{swechat}, the median per-session average context length is about 62k tokens per request, and the median session length is 38 requests.
As each session progresses, agents will frequently edit this context for various reasons, including: 1) to prevent accuracy degradation~\cite{lostmiddle, contextrot}, 2) to remove errors made previously during the session~\cite{lostmultiturn,selfconditioning}, 3) to reduce cost by masking or summarizing old context~\cite{sweagent,complexitytrap,cliffcompaction}, and 4) as part of memory agent architectures~\cite{memgpt}.
Serving such long sessions at scale requires infrastructure tailored to the characteristics of LLM inference.
Modern serving systems often rely on prefill-decode (P/D) disaggregation, which is supported by major engines such as vLLM and SGLang.

Unfortunately, such context edits can introduce significant performance and monetary costs because they eliminate much of the benefit from prefix caching.
In the absence of edits, the serving engine could maintain the KV cache of the session context, requiring prefilling only for the newly appended tokens \cite{vllm, sglang, cachedattention}.
Editing earlier context breaks this by invalidating the KV cache for the suffix, forcing the suffix to be prefilled again prior to generation (i.e., the \emph{miss penalty}).
For some classes of edits, such as those that mask or summarize old context, this can lead to invalidation of most (if not all) of the KV cache.
Each miss delays the agent's next action, and repeated misses accumulate into longer task completion times.
In this paper, we focus on the problem of the performance cost as a result of the miss penalty while ensuring lossless output quality.

Addressing the miss penalty is challenging because of the critical path for finishing prefill before starting decoding.
P/D disaggregation can help by shifting all prefill work to other GPUs; however, this only protects performance observed by \emph{other} sessions that are batched with the session that encountered the cache miss and does not help the affected session.
Selective recompute, which works well in RAG scenarios with short LLM output, cannot work well in long session agentic workflow~\cite{contextpilot,agentkvshift}.
Compacted KV captures the basic information of the context but leads to quality loss in the output~\cite{scbench}.

We propose \sys\footnote{\emph{\sys} is Italian and Spanish for callus. In biology, callus is provisional tissue that seals a plant wound or bridges a bone fracture, letting the part keep working as it heals before regular tissue replaces it.}, a system that reduces the miss penalty in a lossless manner through selective speculative execution.
Our key insight is that the miss penalty is a wait on the decode instance, and that instance already holds the request's \invkv, which is close enough to the exact KV to draft from.
\sys breaks up the critical path by allowing speculative decoding to proceed with the target model but on the (repositioned) invalidated KV cache while prefilling takes place for the invalidated suffix.
\sys streams the updates from prefilling and modifies the KV cache for decoding in-place as the drafting proceeds.
Once the complete KV cache arrives, \sys verifies the draft tokens and commits the accepted prefix before resuming (standard) decoding.
\sys manages decode resources by scheduling speculations during periods of underutilization while preempting speculations when tokens are likely to not be accepted.
This allows \sys to significantly reduce the miss penalty while gracefully degrading to non-speculation performance under saturated load at the decode instances.%

We make the following contributions:
\begin{enumerate}[leftmargin=*,itemsep=0pt,topsep=0pt]

\item We establish invalidated KV as a high-quality draft source and characterize the error and recovery patterns that guide \sys's design.

\item We design \sys, including several techniques to improve performance for single speculations (e.g., decoding while prefill streams KV cache updates) as well as across speculations (e.g., determining when to yield a speculation based on its cumulative acceptance probability).

\item We implement \sys on disaggregated vLLM with live cross-instance KV transfer, without changing vLLM's source or kernels. We evaluate \sys on SWE-agent workloads across three model families. \sys achieves 1.06--1.24$\times$ miss-request speedups over \base at moderate load on our small, real-world testbed. Our cluster-scale simulations demonstrate its benefits extend across varying P:D allocations.

\end{enumerate}

\begin{figure}[t]
    \centering
    \includegraphics[width=\linewidth]{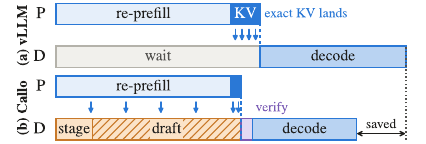}
    \caption{One edited request. (a)~\base, like other conventional systems supporting P/D, waits on the D-instance for the exact KV. (b)~\sys{} drafts on \invkv meanwhile, then verifies. %
    }
    \label{fig:design-timeline}
\end{figure}

\section{Background}
\label{sec:background}

\subsection{Context Edits and Prefix Caching}
\label{sec:bg-miss}

Prefix caches reuse KV only for an exactly matching prefix~\cite{vllm, sglang}.
An edit changes later tokens' preceding context or positions, so their cached KV no longer matches the new prompt even when their token IDs are unchanged~\cite{patchkv, manus}.
This \emph{edit-induced cache miss} requires computing the suffix again, a \emph{re-prefill}, to obtain \emph{exact KV}.
Later requests of the session build on the edited prompt, so the suffix's old KV will not match them either~\cite{sweagentcode}.
The edit has effectively invalidated this KV, which we call the \emph{\invkv}, and it remains cached until discarded when its memory is needed.

Removing redundant content cuts input tokens by 40--60\% without reducing solve rates~\cite{agentdiet}, yet the cost of re-prefill discourages agent systems from doing so~\cite{manus,sweagentcode,goose}.
Today's append-only defaults thus reflect what prefix caching makes affordable, not what agents need.

\subsection{Prefill/Decode Disaggregation}
\label{sec:bg-pd}

Long prefill steps can delay ongoing decoding when both phases share GPUs~\cite{distserve}.
Their resource demands also differ: prefill is compute-intensive, while decoding is memory-bandwidth-intensive~\cite{splitwise}.
Prefill/decode (P/D) disaggregation places the phases on separate GPUs to avoid this interference and allow independent resource allocation and hardware selection.
A \emph{prefill instance} (P-instance) computes and transfers KV to a \emph{decode instance} (D-instance), which generates the response.
While the P-instance re-prefills an edited request, the D-instance can continue serving other requests.

\subsection{Speculative Decoding}
\label{sec:bg-spec}

In speculative decoding~\cite{leviathan, specsampling}, a drafter proposes tokens and the target model verifies them in one forward pass.
Greedy verification accepts the longest prefix matching the target's argmax, then adds a correction token at the first mismatch or a bonus token if all drafts match.
This \emph{\fixedtie} rule preserves the target's choice, and when target logits tie, the lowest token ID wins.
An \emph{\anymax} variant also accepts any draft tied exactly for the maximum target logit.
Under sampling, probabilistic acceptance and correction preserve the target distribution.
In either case, drafts may use approximate state. Draft quality affects acceptance, while verification
preserves the chosen decoding rule.

\section{Motivation}
\label{sec:motivation}

\begin{figure*}[t]
\centering
\includegraphics[width=\textwidth]{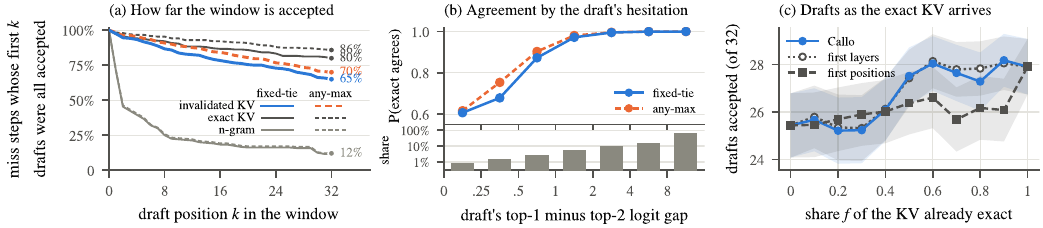}
\caption{Drafts made on \invkv (Llama-3.3-70B-Instruct FP8, SWE-agent L5, 32-token window, offline). (a)~Miss steps
whose first $k$ drafts are all accepted (240 steps), against drafts on the exact KV and n-gram drafts. (b)~Agreement by
the draft's top-1 minus top-2 logit gap (1{,}400 steps); below, the share of positions in each bin. (c)~Accepted drafts when a share $f$
of the \invkv's position--layer cells is replaced by exact KV before drafting: \sys's delivery order and two reference
orders (240 steps; bands: 95\% confidence intervals).
}
\label{fig:acceptance}
\end{figure*}

\subsection{Cache Miss as Opportunity for Speculation}
\label{sec:opportunity}

An edit-induced miss puts exact re-prefill on the request's critical path, even when the previous request's KV is still cached on the D-instance.
We quantify this opportunity on SWE-bench Verified sessions recorded under edit policies from prior work, rather than cache-constrained append-only defaults (\S\ref{sec:setup}).
Our main workload follows SWE-agent's published configuration: it retains the last five tool outputs in full and replaces older ones with placeholders~\cite{sweagent}, but we also evaluate retaining ten~\cite{complexitytrap}.

We present the results in Table~\ref{tab:workloads}.
On Llama-3.3-70B, 79\% of requests in the main workload result in edit-induced misses.
A miss prefills 3.1k tokens at the median.

\begin{table}[t]
\centering
\caption{Workloads recorded on SWE-bench Verified and replayed in the evaluation (\S\ref{sec:setup}). L$n$ keeps the last $n$ observations in full. For each miss request, Prefill counts tokens computed by P-instances, and Reusable counts those whose \invkv remains on D-instances.}
\label{tab:workloads}
\vskip 0.1in
\small\setlength{\tabcolsep}{4pt}
\begin{tabular}{@{}lrrrrr@{}}
\toprule
 & & & & \multicolumn{2}{c}{Per miss request} \\
\cmidrule(lr){5-6}
Workload & Prompt & Reply & Miss & Prefill & Reusable \\
 & mean & mean & rate & p50/p90 & p50/p90 \\
\midrule
Llama L5 & 13.5k & 117 & 79\% & 3.1k/8.9k & 2.5k/7.4k \\
Llama L10 & 15.3k & 121 & 58\% & 8.3k/18.3k & 7.6k/16.8k \\
\midrule
Gemma L5 & 15.4k & 199 & 74\% & 5.2k/8.3k & 4.2k/7.2k \\
Gemma L10 & 17.5k & 189 & 53\% & 10.9k/15.5k & 9.8k/14.5k \\
\midrule
Mistral L5 & 18.1k & 173 & 87\% & 4.7k/14.1k & 3.9k/13.4k \\
Mistral L10 & 19.7k & 224 & 73\% & 8.4k/15.9k & 7.8k/15.3k \\
\bottomrule
\end{tabular}

\vskip -0.1in
\end{table}

Directly decoding on approximate KV can change outputs, and early deviations can propagate across an agent session~\cite{vericache, lostmultiturn}.
\sys therefore commits tokens only on exact KV. Drafting, however, need not wait.
While the P-instance recomputes the exact KV, the D-instance can use its spare capacity (\S\ref{sec:bg-pd}) to draft with the target model on the \invkv.
Each accepted draft token removes a decode step from the request's critical path.

\subsection{Invalidated KV as Speculative State}
\label{sec:acceptance}

\begin{figure*}[t]
    \centering
    \includegraphics[width=\textwidth]{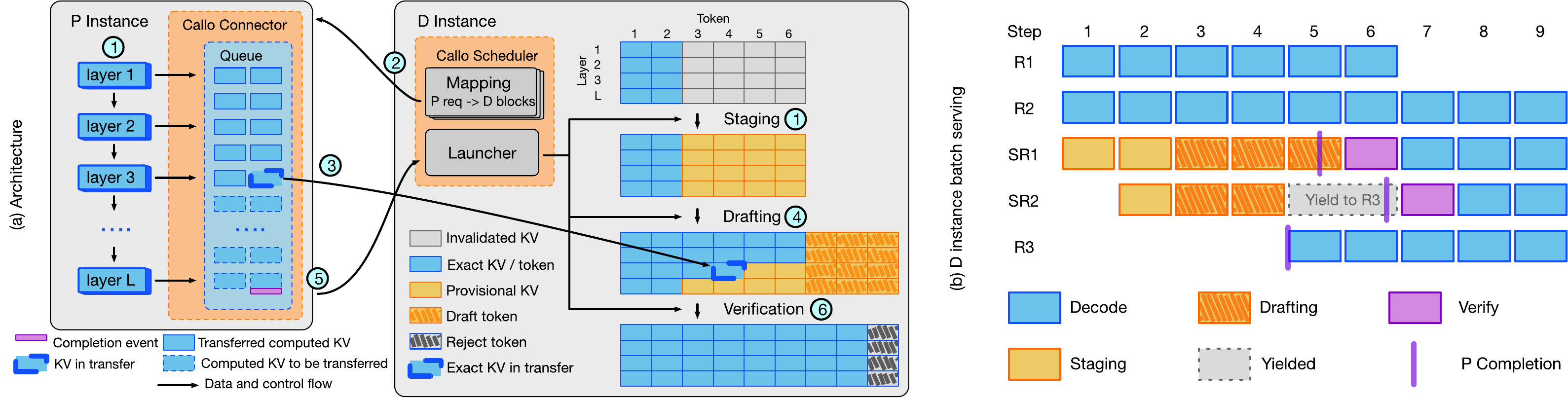}
    \caption{Architecture of \sys and the life of a miss request. (a)~Numbers give the order of events in this example: (1)~the P-instance starts the re-prefill while the D-instance starts staging, (2)~the D-instance registers the request's blocks with the P-instance, (3)~the P-instance's connector starts writing each finished layer into them, (4)~the D-instance starts drafting, (5)~the P-instance signals completion, and (6)~the D-instance starts verification. (b)~The D-instance's batch with a limit of four requests, of which SR1 and SR2 speculate: when R3 arrives, SR2 yields its slot but keeps its drafts for verification.}
    \label{fig:design-kv-states}
\end{figure*}

The same dependence on earlier tokens that invalidates the KV after an edit also keeps it close to the exact KV.
A retained token is unchanged, and re-rotation adjusts its keys to their new positions. An edit invalidates its KV by changing the context it attends to, most of which remains unchanged (Appendix~\ref{app:per-layer}).
At the output the difference is small.
In our offline replay (Appendix~\ref{app:offline}), we feed the model on \invkv the tokens that exact decoding generates.
The predicted next token matches that on the exact KV at 98.3\% of positions, and the next-token distribution is within 0.01 total variation of the exact one at 80\%.
\Invkv is thus not noise: invalid as committed state, yet close enough to serve as speculative state. Nor is it an approximation in the usual sense.
Unlike compressed, quantized or sparsified KV, which approximates the current context~\cite{quantspec, magicdec, triforce}, \invkv misses only the edit.
Its error therefore has a structure: where it enters, which tokens it flips, and how it heals as the exact KV arrives.
The rest of this section measures this structure, and \sys's design rests on it (Table~\ref{tab:findings}).

\begin{table}[t]
\centering
\caption{Properties of \invkv measured in \S\ref{sec:acceptance}, and the design choices of \sys they motivate.}
\label{tab:findings}
\vskip 0.1in
\footnotesize
\begin{tabular}{@{}ll@{}}
\toprule
Property of \invkv & Design choice \\
\midrule
Close to the exact KV & Inherit and reposition (\S\ref{sec:staging}) \\
Near-exact acceptance (Fig.~\ref{fig:acceptance}a) & Target-model drafting (\S\ref{sec:overview}) \\
Errors at near-ties (Fig.~\ref{fig:acceptance}b) & Cumulative early stop (\S\ref{sec:policy}) \\
Heals layer by layer (Fig.~\ref{fig:acceptance}c) & Layer-wise streaming (\S\ref{sec:streaming}) \\
Mixing never hurts (Fig.~\ref{fig:acceptance}c) & Lock-free drafting (\S\ref{sec:streaming}) \\
\bottomrule
\end{tabular}
\vskip -0.1in
\end{table}

Drafts made on \invkv survive exact verification almost as often as tokens decoded directly on the exact KV.
We replay miss requests of our main workload offline and draft 32 tokens on \invkv re-positioned as in \S\ref{sec:staging}. The whole window is accepted in 65\% of miss requests under \fixedtie verification and 70\% under \anymax verification.
Tokens decoded on the exact KV reach 80\% and 86\%, a ceiling below 100\% because single-token decoding and the multi-token verification pass differ numerically at near-ties.
Prompt-lookup n-grams~\cite{pld}, which copy continuations of matching prompt sequences, need no KV but reach only 12\% (Figure~\ref{fig:acceptance}a).
The target model only reaches 10\% when drafting on the valid prefix and newly added tokens, with the retained suffix left out.
Drafts on \invkv thus need both the target model and the context that \invkv holds.
Appendix~\ref{app:accept-models} gives the same measurements for the other two models.

Drafts fail where the model itself hesitates (Figure~\ref{fig:acceptance}b).
Where the draft's top-1 logit exceeds its top-2 by at least one, the exact model agrees at 97--100\% of positions; where the gap is below 0.5, at only 65\%.
These near-ties account for 2.4\% of draft positions but 60\% of first mismatches. Gaps below one account for 5\% of positions and 85\% of first mismatches.
\Invkv changes token choices mainly at near-ties, and the model is rarely confident in drafts that later fail verification.
Where a draft will fail is therefore foreseeable while it is being made, in its own logit gap.

Drafting need not wait for all of the exact KV and can  use each part as the P-instance streams it in (Figure~\ref{fig:acceptance}c).
A mixture of invalidated and exact KV is a nonstandard state, yet at no share does it accept significantly fewer drafts than the \invkv alone, and some mixtures accept significantly more.
What the exact KV adds depends on the order in which it arrives.
When whole layers turn exact first, acceptance stays flat while the first layers arrive and then rises.
Once half of the layers are exact, drafting accepts significantly more drafts than on \invkv and as many as on the exact KV.
The same shares delivered position by position, as fine-grained chunked prefill completes them, give no significant gain over \invkv and stay below the exact KV.
\sys follows a third order, pushing each 8k-token prefill chunk layer by layer in order to keep vLLM unchanged. Its acceptance closely tracks the first-layers reference.
The state built from \invkv thus heals in place as exact KV arrives. Layer-wise delivery makes better intermediate states available for drafting.

\section{Design}
\label{sec:design}

\subsection{Overview}
\label{sec:overview}

\sys breaks the serial prefill--decode critical path of an edit-induced cache miss by overlapping exact re-prefill with speculative decoding, where the \invkv serves as a good source for drafts (Figure~\ref{fig:design-timeline}).

Our design follows three principles:
\begin{enumerate}[leftmargin=*,itemsep=0pt,topsep=0pt]
    \item \label{pr:lossless}\emph{Lossless:} \sys uses \invkv for proposals, but only commits based on the exact KV.

    \item \label{pr:workload}\emph{Harness Agnostic:} \sys finds every edit itself and does not require modifying the harness.

    \item \label{pr:spare}\emph{Prioritize Exact:} \sys never allows speculation to block decoding on exact KV and terminates speculations early when they degrade (i.e., unlikely to verify subsequent tokens).

\end{enumerate}

Figure~\ref{fig:design-kv-states}a shows \sys's components and the phases of one miss request.
When the \sys scheduler on the D-instance decides to admit the speculation for a request, it first \emph{stages} the request, building its \emph{\provkv} by repositioning invalidated KV and computing KV for new tokens (\S\ref{sec:staging}).
The D-instance then \emph{drafts} with the target model on \provkv while \sys's connector on the P-instance writes each layer's exact KV into place as soon as it is computed (\S\ref{sec:streaming}).
The P-instance's completion event starts \emph{verification}, which commits the accepted drafts before ordinary decoding resumes (\S\ref{sec:verify}).
Speculating requests share the D-instance's batch with ordinary requests and yield to them (Figure~\ref{fig:design-kv-states}b, \S\ref{sec:policy}); requests that cannot speculate follow the ordinary path.

\subsection{Building the Provisional KV}
\label{sec:staging}

The \invkv comes from the previous request of the same session. When that request finishes, its KV stays in the decode instance's prefix cache, like that of any finished request, until its memory is needed. \sys builds its provisional state by inheriting the KV of retained tokens from this cache and repositioning it for the edited prompt.

The prefix up to the edit is reused exactly, through the engine's ordinary prefix cache.
The suffix is more challenging: even when the tokens are unchanged, the edit can shift their positions making the cached KV no longer line up with the new prompt.
\sys therefore aligns the previous request's tokens, including those of its reply, with the new prompt. It finds runs of identical tokens and keeps them in order, so that each position of the new prompt either maps to a position of the previous request or is new. It then copies the KV of the mapped positions into the request's blocks at their new positions, re-rotating each key with a fused Triton kernel. Once the copy completes, the decode instance releases its temporary references to the source blocks, allowing ordinary prefix-cache eviction when no other references remain.
The copy is the only part of staging that runs outside a D-instance batch.

At this point, the remaining positions have no counterpart in the previous request.
They are the spans that the edit introduced in the middle of the context and the \emph{tail}, the newly appended messages at its end.
The decode instance computes them as the request's own prefill chunks in its ordinary batch, alongside other requests' decode rows, and keeps them apart from the blocks that the prefill instance writes, so that the prefill instance can write the request's blocks without any lock (\S\ref{sec:streaming}).
Each chunk holds at most $C_s=112$ tokens of one span, so a long span or several separate spans take several steps instead of stalling the batch. The last chunk ends with the prompt's last token, whose output is the first draft.
On the main workload (\S\ref{sec:setup}), staging computes a median of 424 tokens per speculating miss request, almost all from the newly appended turn.
Since staging delays every request in the D-instance batch, \sys applies a policy to determine when to admit speculations based on the number of staging tokens and the active requests in the batch (see \S\ref{sec:policy}).

Hybrid models need no new mechanism. Gemma-4-31B interleaves full-attention layers with sliding-window layers that read only the previous 1{,}024 positions, and vLLM keeps a sliding-window layer's KV only for that last window. \sys stages each of vLLM's KV cache groups on its own, as in a dense model, without changing vLLM's hybrid KV cache manager or its kernels.

\subsection{Overlapping Drafting with Exact Re-prefill}
\label{sec:streaming}

\sys does not change the computation of the re-prefill, but it does push the KV to the D-instance at a finer granularity.
As in vLLM's push mode, the D-instance tells the P-instance where the request's blocks are in a \emph{registration}.
After each attention layer, the prefill instance records a CUDA event and hands the layer's blocks to a writer thread, which issues the push once the event has completed and the registration has arrived. The prefill instance can hence recompute the suffix while the decode instance is still staging. A final push carries the completion notification and is issued only after every layer's push has completed. If \sys admits the speculation, it will keep drafting until this notification arrives.
All of the operations, prior to verification, are asynchronous: the P-instance forward pass does not block waiting on the D-instance registration, and D-instance drafting does not block waiting on the P-instance to start prefill.

There are two key aspects to our design that help reduce the end-to-end latency.
First, streaming each layer's KV as it is produced spreads transfers over prefill, reducing bursts at the end (especially when requests share a link).
Second, layer-major KV updates also support
higher draft acceptance than position-major updates
(Table~\ref{tab:findings}).
Ideally, this means delivering one layer's exact KV for the entire suffix before moving to the next layer. To preserve vLLM's chunked prefill, \sys instead streams layer by layer within each chunk. On our GPUs, vLLM's default chunk holds up to 8{,}192 tokens, and 88\% of re-prefills fit in one, so most miss requests still receive their exact KV in pure layer order.

The D-instance drafting will read the KV blocks while the prefill instance writes into them, without any lock or synchronization. While drafting overlaps with the P-instance's writes, the two instances write to separate blocks. The decode instance sends the registration only after repositioning the retained KV, so incoming exact KV always overwrites the provisional copy in-place. It keeps the prompt KV it computes during staging in \emph{scratch blocks}, separate from the blocks that receive the prefill instance's writes (Appendix~\ref{app:impl-scratch}). Staging and drafting read these scratch blocks together with the shared context that the prefill instance refreshes in place.
The draft will write past the end of the prompt, but this is not an issue because the P-instance will never write to those locations.
D-instance reads may interleave with P-instance writes, since a draft may read a row before, after or while its exact version lands. This race affects only which tokens are proposed; however, our mixed-state experiments in \S\ref{sec:acceptance} show no significant loss in draft acceptance relative to \invkv alone, and higher acceptance once half of the layers are exact.
At verification, the D-instance switches to the request's own blocks
and frees the scratch blocks. They take a small share of the KV pool (\S\ref{sec:memory}) and are capped (Appendix~\ref{app:policy}).

\subsection{Verification}
\label{sec:verify}

Verification begins only after the completion notification (\S\ref{sec:streaming}) confirms that the request's entire prompt KV is exact and available in the D-instance memory.
If the D-instance did not admit the speculation for a request, or if the exact KV arrives before the first draft, there is nothing to verify; however, in other cases, there are one or more draft tokens produced that need to be verified.
In the request's next scheduled step, the decode instance runs one forward pass over the last prompt token and all $k$ drafts, as a $(k{+}1)$-token chunk in its ordinary batch.
The verification pass reads the exact KV and recomputes the drafts’ own KV. Its logits therefore depend only on exact KV, regardless of how the drafts were generated (principle~\ref{pr:lossless}).
The verifier accepts draft tokens up to the first rejection and discards the remaining tokens.
It commits the accepted prefix together with a correction token, or a bonus token if all drafts are accepted (\S\ref{sec:bg-spec}). The accepted prefix retains the KV recomputed during verification, so no further repair is needed.
\sys verifies with \fixedtie decoding by default, which maintains lossless behavior, but also supports \anymax verification, which accepts any co-maximum of the logits, as well as speculative sampling~\cite{leviathan, specsampling}.

\subsection{Decode-Side Policy}
\label{sec:policy}

\sys uses spare decode capacity to advance requests while their exact KV is being computed and transferred. Relaxing the prefill--decode dependency gives the scheduler more work to choose from, but offers less benefit when ordinary requests already keep the decode instance busy. The policy therefore gives ordinary requests priority and backs off as they compete for capacity (principle~\ref{pr:spare}). Three rules implement this policy.

\emph{Admission:} A miss request needs a free slot that no ready request needs.
The required cached KV must include at least 256 reusable tokens of \invkv,
and the request's staging tokens $s$, multiplied by the number $B$ of requests in the batch including itself, must stay within a budget $K$, which bounds the delay that staging adds to every request in the batch.
We only decide whether to admit a speculation immediately when a request arrives and will not queue speculations.

\emph{Early Stop:} For greedy decoding, \sys estimates the probability that all drafts so far survive verification as the product of per-draft agreement probabilities, calibrated offline from each model's logit gaps (Table~\ref{tab:findings}).
We call this the \emph{cumulative acceptance probability}. \sys stops drafting when it falls below $\theta = 0.3$, but this is a configurable parameter.

\emph{Yielding:} When ready requests need more slots than are free, drafting requests give theirs up, those with the lowest cumulative acceptance probability first. They keep their KV and drafts for verification, as SR2 does in Figure~\ref{fig:design-kv-states}b.

Our default uses the same thresholds across serving configurations, with a confidence mapping calibrated per model. Appendix~\ref{app:policy} details parameter choices and extensions using richer model or workload information.

\subsection{Serving Integration}
\label{sec:integration}

\sys loads through vLLM's scheduler and KV-connector extension points (the orange components in Figure~\ref{fig:design-kv-states}a), without changing its source or existing kernels.
Staging, drafting and verification use vLLM's existing batched execution path. KV written during speculation is kept out of the prefix cache to prevent other requests from reusing unverified state.
If speculation is abandoned or preempted, the request returns to ordinary service while retaining blocks that the P-instance may still write. Appendix~\ref{app:impl} details the integration and block lifetimes.

\section{Evaluation}
\label{sec:evaluation}

We evaluate how \sys's latency gains vary with load and available decode capacity, and separate the contributions of speculation and its transfer path. We also measure memory use and interference with ordinary requests, and test the system at cluster scale.

\subsection{Experimental Setup}
\label{sec:setup}

\noindent\textbf{Hardware and models.}\quad
One P-instance and one D-instance each run on
an NVIDIA H200 NVL (141\,GB), connected by NVLink.
The instances run vLLM 0.29~\cite{vllm}, transferring
KV through NIXL.
We evaluate Llama-3.3-70B-Instruct~\cite{llama3} in FP8,
and Gemma-4-31B~\cite{gemma4} and
Mistral-Small-3.2-24B~\cite{mistralsmall} in bf16.
Unless stated otherwise, results use \mainmodel on \mainwl.

\noindent\textbf{Workloads.}\quad
We use SWE-agent~1.1~\cite{sweagent} with the published configuration: it retains the last five tool outputs in full (replacing older ones with placeholders), but we also evaluate retaining ten~\cite{complexitytrap}.
For each model and setting, we record prompts, replies
and tool times on the same 100 SWE-bench
Verified~\cite{swebench} issues at temperature 0.
We held out 20 issues to calibrate early stopping.
We evaluate only the remaining eligible sessions
(Appendix~\ref{app:workload}).
Sessions arrive at a fixed rate on a seeded schedule
shared by all systems. Each system replays the recorded
prompts and generates the recorded number of reply tokens.
Each session waits for its reply and the recorded tool
time before issuing the next request.
We measure only sessions that arrive after a warm-up, and further sessions keep arriving until these finish, so every measured request runs at the target load (Appendix~\ref{app:replay}).

\noindent\textbf{Systems.}\quad
The baseline is unmodified \base with native NIXL push, using the same version as \sys.
Its native connector sends a request's KV only after its prefill finishes.
\sys includes its complete transfer and speculation paths
and uses \fixedtie verification.
Its default policy uses $\theta=0.3$, at least 256 reusable
tokens of \invkv, and a scratch cap of 10\% of the D-instance's KV pool.
The staging budget $K$ is 4{,}096 token-requests (\S\ref{sec:policy}).
We also evaluate \sysstream, which keeps \sys's transfer path but disables speculation, to separate the two contributions.

\noindent\textbf{Metrics.}\quad
We report the ratio of \base's mean request latency to
\sys's on the same requests, separately for all requests
and misses.
We obtain 95\% confidence intervals by resampling the measured sessions.
A request's baseline miss penalty is the wait from arrival
until exact KV lands on the D-instance.
Its \emph{folded share} is the net reduction in request
latency divided by this penalty, accounting for staging
and verification.

\noindent\textbf{Load.}\quad Near capacity, the system keeps pace with arrivals,
but suffers from queueing latency and has little headroom.
We estimate capacity by sweeping \base's session arrival rate
(Appendix~\ref{app:capacity}). Load levels come from a week of Azure's coding traffic~\cite{dynamollm}.
We take its 99th-percentile request rate across 10-minute windows as
the provisioning peak. The 25th and 75th percentiles are approximately
12\% and 52\% of this peak. We therefore evaluate each configuration
at \emph{light load} (12\% of capacity), \emph{moderate load} (52\%)
and \emph{full load} (100\%).

\subsection{End-to-end serving}
\label{sec:e2e}

\begin{table}[t]
\centering
\caption{Mean request-latency ratios (\base/\sys) for miss requests
and all requests. D-instance batch limit: 16. Ratios use requests completed by both systems.}
\label{tab:main}
\vskip 0.1in
\small
\setlength{\tabcolsep}{5pt}
\begin{tabular}{@{}lrrrrrr@{}}
\toprule
 & \multicolumn{2}{c}{\shortstack{\strut Light\\(12\%)}} & \multicolumn{2}{c}{\shortstack{\strut Moderate\\(52\%)}} & \multicolumn{2}{c}{\shortstack{\strut Full\\(100\%)}} \\
\cmidrule(lr){2-3}\cmidrule(lr){4-5}\cmidrule(lr){6-7}
Workload & Miss & All & Miss & All & Miss & All \\
\midrule
Llama L5 & 1.13 & 1.10 & 1.19 & 1.14 & 1.32 & 1.27 \\
Llama L10 & 1.24 & 1.17 & 1.24 & 1.19 & 1.21 & 1.17 \\
\midrule
Gemma L5 & 1.07 & 1.05 & 1.06 & 1.05 & 1.10 & 1.09 \\
Gemma L10 & 1.11 & 1.07 & 1.11 & 1.08 & 1.12 & 1.10 \\
\midrule
Mistral L5 & 1.15 & 1.14 & 1.16 & 1.14 & 1.16 & 1.15 \\
Mistral L10 & 1.21 & 1.15 & 1.19 & 1.14 & 1.15 & 1.13 \\
\bottomrule
\end{tabular}

\vskip -0.1in
\end{table}

Across the evaluated models and context-management settings, \sys reduces mean request latency at both light and moderate load (Table~\ref{tab:main}). On the main configuration, all requests are 1.10$\times$ and 1.14$\times$ as fast as with \base, respectively. How these gains change as load rises depends on which instance becomes the bottleneck.

\begin{figure}[t]
\centering
\includegraphics[width=\columnwidth]{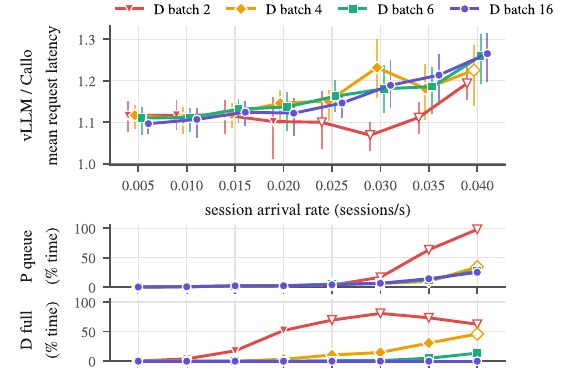}
\caption{All-request speedup (\base/\sys) across D-instance batch limits.
Lower panels show baseline time with requests queued at the P-instance
and with the D-instance full while requests wait. Error bars: 95\% CIs.
Hollow markers: either run left measured requests unissued
or unfinished.}
\label{fig:D_bottleneck}
\end{figure}

Figure~\ref{fig:D_bottleneck} sweeps the arrival rate on the main workload with D-instance batch limits of 16, 6, 4 and 2.
Reducing the batch limit brings decode contention at lower arrival rates.
With a limit of 16, the all-request speedup over \base rises from 1.10$\times$ at light load to 1.27$\times$ at full load.
The P-instance becomes the bottleneck while the D-instance still has free slots, and speculation remains effective as longer waits for the P-instance leave the D-instance more time to draft.
High overall load can therefore coexist with spare decode capacity.
Limits of 6 and 4 follow the same curve, since their D-instance fills only near capacity.
With a limit of two, the D-instance runs full about half the time at its measured capacity of 0.02 sessions/s, and the speedup stays around 1.10--1.12$\times$.

Speculation alone would not produce these trends.
Since \sys backs off as the D-instance fills (\S\ref{sec:policy}), its gain should hold with a limit of 16 and shrink with a limit of two.
The speedup instead grows with a limit of 16 and holds with a limit of two, because \sys's transfer path saves more as the P-instance becomes busier.

\begin{figure}[t]
\centering
\includegraphics[width=\columnwidth]{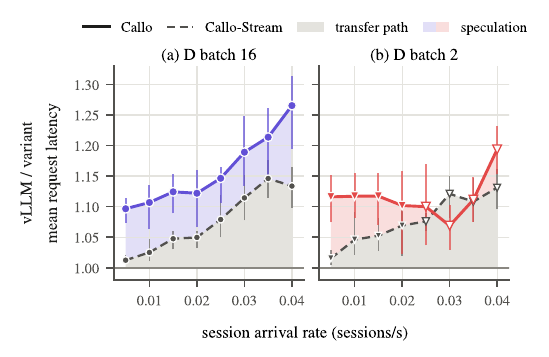}
\caption{Contributions of streaming and speculation to
all-request speedup. Dashed: measured \sysstream.
Both curves use requests completed by all three systems.
CIs and hollow markers follow Figure~\ref{fig:D_bottleneck}.}
\label{fig:D_split}
\end{figure}

Figure~\ref{fig:D_split} confirms this split.
Relative to \sysstream, speculation reduces mean all-request latency by 8\% at light load and 10\% at full load with a D-instance batch limit of 16.
Its contribution narrows as decode resources become contended, as the policy intends.
With a limit of two, speculation reduces latency by 3\% at capacity, and the transfer path accounts for most of the gain. Past capacity (hollow markers), queues keep growing, so these ratios are not steady-state gains (Appendix~\ref{app:beyond-capacity}).

\begin{figure}[t]
\centering
\includegraphics[width=\columnwidth]{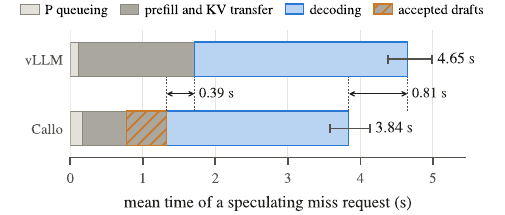}
\caption{Mean latency breakdown for 624 miss requests that speculate
under \sys at moderate load. Whiskers: 95\% CIs.
Grey ends at exact-KV arrival, estimated for \base.
Hatched: accepted drafts, in equivalent baseline
decode time.}
\label{fig:micro}
\end{figure}

Figure~\ref{fig:micro} shows how \sys shortens the critical path of speculating miss requests at moderate load.
Under \base, these requests wait 1.7\,s for exact KV and then spend 2.9\,s decoding.
With \sys, exact KV arrives 0.4\,s sooner, and the time remaining
after its arrival falls by another 0.4\,s.
Accepted drafts advance the reply during the wait, leaving less decoding after exact KV is ready.
Together, the savings shorten the request by 0.8\,s on average, folding 47\% of its miss penalty.

The context-management setting also changes the opportunity per
miss.
Across all three models, retaining ten observations produces
fewer misses but longer suffixes to prefill
(Table~\ref{tab:workloads}).
On \mainmodel, the miss rate falls
from 79\% to 58\%, while the median prefill per miss grows from
3.1k to 8.3k tokens. At moderate load, the miss-request gain rises
from 1.19$\times$ to 1.24$\times$, and the all-request gain from
1.14$\times$ to 1.19$\times$ (Table~\ref{tab:main}).
A lower miss rate can thus coexist with a larger end-to-end gain when each miss exposes more work to overlap.

\subsection{Memory}
\label{sec:memory}

\sys stays within \base's GPU memory budget by setting aside 0.75\,GiB on the D-instance for runtime buffers.
This reduces the D-instance's KV pool by 1.5\% on the main configuration.
Across the evaluated runs, recorded allocator reservations beyond the startup footprint use at most 0.29\,GiB of this headroom.

\Invkv stays in the prefix cache and needs no extra reservation (Table~\ref{tab:admission}).

At moderate load on the main configuration, scratch blocks occupy 0.24\% of the D-instance's KV pool on average and 2.71\% at peak.
Requests that make progress while waiting also finish and release their KV sooner.
Overall, mean KV pool occupancy is 24.6\% under \sys, compared with 27.2\% under \base.
Peak occupancy is similar, at 75.4\% and 75.1\%, respectively.

\subsection{Cluster-scale simulation}
\label{sec:sim}

To examine how \sys behaves beyond a single P/D pair, we simulate a 96-GPU cluster serving \mainmodel on \mainwl.
We evaluate P:D ratios of 1:2, 1:1, 2:1 and 3:1 with a cache-aware router that preserves session affinity while balancing load.
For each P:D allocation, we sweep the arrival rate, taking \base's capacity as 100\% load.
Engine-step times are fitted to testbed measurements, and draft acceptance is modeled from recorded speculations.
Appendix~\ref{app:sim} describes the model and its calibration against the testbed.

\begin{figure}[t]
\centering
\includegraphics[width=\columnwidth]{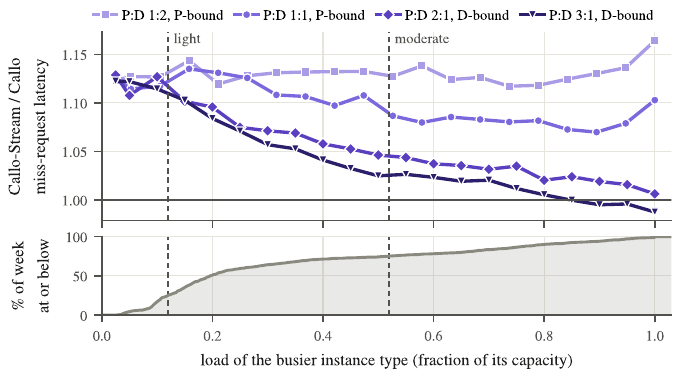}
\caption{Miss-request speedup over \sysstream across P:D allocations
in a 96-GPU cluster.
Load is relative to each allocation's \base capacity.
Below: share of Azure's week at or below each load; the dashed light (12\%) and moderate (52\%) loads are its 25th
and 75th percentiles.}
\label{fig:sim}
\end{figure}

Figure~\ref{fig:sim} compares \sys with \sysstream to isolate speculation's contribution.
At light load, all four P:D allocations achieve similar miss-request speedups of 1.110--1.131$\times$.
As load rises, configurations with fewer D-instances have less room for speculation, widening the speedup range to 1.025--1.128$\times$ at moderate load.
At 2:1, the gain persists up to capacity. At 3:1, the ratio returns to 0.99--1.00 beyond about 85\% of capacity.

The Azure trace spends most of the week at or below moderate load (lower panel), where all four allocations benefit from speculation.
Weighting the load-sweep results by the week's traffic gives miss-request speedups of 1.130$\times$, 1.095$\times$, 1.044$\times$ and 1.027$\times$ for P:D ratios of 1:2, 1:1, 2:1 and 3:1, respectively.

\subsection{Correctness and Overheads}
\label{sec:correctness}

We validate correctness in instrumented \fixedtie runs,
separate from the timed experiments.
We compare fingerprints of the prompt KV read during
verification with those of its exact sources.
We check acceptance decisions and committed tokens against
the target model's argmax from the same verification pass.
Outputs may differ across runs because batch composition
can affect numerical results. We also check the state after rollback.
Across 1{,}928 verification events and 44{,}792 committed
tokens, we find no KV-fingerprint mismatches and no
acceptance, output, or rollback-state violations.

Staging, drafting and verification share the D-instance's batches with ordinary decoding.
We compare requests that do not speculate under \sys with the same requests under \sysstream.
On the main configuration, their mean time from the P-instance's response to reply completion is about 2\% longer at low load and 6\% longer near capacity.
The end-to-end gains in \S\ref{sec:e2e} include this cost.

\section{Ablation Study}
\label{sec:ablation}

On the testbed, we run one request at a time with both
instances initially idle. To isolate the mechanism,
these runs disable admission thresholds and
confidence early stopping. Staging is unchunked, and
the candidate window is capped at 111 tokens.
We then use the cluster simulator to study streaming
and decode-side policy under resource contention.
Additional results and more detailed analyses appear in Appendices~\ref{app:policy} and~\ref{app:sim}.

\begin{table}[t]
\centering
\caption{Single-request ablations on \mainmodel and \mainwl.
Folded: net share of miss penalty hidden.
Speedup: ratio of mean miss-request latencies (\base/variant).
Accepted: mean accepted drafts per miss request, including zeros.}
\label{tab:ablation}
\vskip 0.1in
\small\setlength{\tabcolsep}{4pt}
\begin{tabular}{@{}lrrr@{}}
\toprule
Variant & Folded & Speedup & Accepted \\
\midrule
\sys (\fixedtie) & 49\% & 1.15 & 22.6 \\
\addlinespace
\quad \sysstream & 4.5\% & 1.01 & -- \\
\quad \anymax verification & 54\% & 1.17 & 24.4 \\
\quad speculative sampling, $T{=}0.8$ & 45\% & 1.14 & 21.4 \\
\addlinespace
\quad n-gram drafts & 18\% & 1.05 & 7.5 \\
\bottomrule
\end{tabular}

\vskip -0.1in
\end{table}

\subsection{Speculation Gains}
\label{sec:ablation:subsec:headroom}

We first measure how much useful decoding can overlap
the wait for exact KV. With \fixedtie verification,
\sys hides 49\% of the estimated miss penalty and
achieves a 1.15$\times$ miss-request speedup over \base.
\sysstream gives a 1.01$\times$ speedup.

Of all drafted tokens, 75\% survive verification,
and 69\% of miss requests accept their entire generated
draft window. In 97\% of these fully accepted windows, drafting ends
when the exact KV arrives.

\subsection{Verification Rules}
\label{sec:ablation:subsec:strict_exact_tie}

\Anymax verification accepts any draft whose target
logit exactly equals the maximum.
It commits 24.4 drafts per miss request and hides
54\% of the miss penalty.

For speculative sampling, both draft and target
distributions use $T=0.8$ and top-$k=64$.
We use the same greedy \base reference and fixed reply
lengths as the other variants.
\sys commits 21.4 drafts per miss request and achieves
a 1.14$\times$ speedup over \base.

\subsection{Draft Source}
\label{sec:ablation:subsec:draft_source}

We replace \invkv-based drafting with prompt-lookup
n-grams to test whether a cheaper draft source can
deliver similar gains. Prompt lookup forms a single
draft window while waiting for exact KV, avoiding
GPU staging and drafting. Both variants use \fixedtie
verification.

Prompt lookup yields 7.5 accepted drafts per miss
request, hiding 18\% of the miss penalty and achieves
a 1.05$\times$ speedup.

\subsection{Transfer Path}
\label{sec:ablation:subsec:streaming_burst}

Layer-wise delivery spreads each request's KV transfer
over prefill, reducing bursts when requests share a link.
Replacing it with asynchronous burst transfer has little
effect in the 1:1 cluster.
In the 3:1 cluster, three prefill instances share each
decode instance's link.
With speculation disabled to isolate transfer costs,
layer-wise delivery raises miss-request speedup from
1.094$\times$ to 1.115$\times$ at 90\% load.
Both variants handle completion promptly.

Most of the transfer-path gain over \base comes from
handling completion without waiting for the P-instance's running
step to end.

\subsection{Decode-Side Policy}
\label{sec:ablation:subsec:policy_study}

Admission balances speculation opportunities against
interference on the D-instance. In the simulated 1:1 cluster, removing the staging
budget barely changes miss-request speedup but eliminates
the gain over \base for non-speculating requests at 90\% load.
Early stopping consistently lowers the fraction of
drafts discarded at verification in every tested P:D
allocation and load. The default $\theta=0.3$ reduces discarded drafts by 30\%
while retaining 97\% of accepted drafts, compared with
disabling early stopping. This benefit also holds at $\theta=0.1$ and $0.5$,
without precise threshold tuning.
Under a fair per-request compute budget, early stopping
improves compute efficiency and creates room for richer
drafting policies, discussed in
Appendix~\ref{app:policy-alternatives}.

\section{Discussion}
\label{sec:discussion}

\sys illustrates a broader use of invalidated caches. A small change to the prefix, positions or model weights can
invalidate a cache without destroying its value as speculative state. Beyond the context edits studied here, the same approach applies when a harness rewrites a
memory block~\cite{memgpt} or refreshes a document in place. Whether a cache computed before an adapter switch also
drafts well remains open.

\sys's saving follows the wait for re-prefill.
Its relative gain depends on how much of a request that wait is.
When requests generate very long outputs, as under extended reasoning, decoding dominates and the saving is only a small fraction of the request.
Latency-sensitive tasks sit at the other end.
They keep reasoning short so that each reply arrives quickly, which makes the re-prefill wait a large part of the latency that remains, and the relative gain rises with it.
\sys is best suited to this type of task.

\sys drafts only during the wait; once ordinary decoding
resumes, conventional speculative drafters~\cite{leviathan, eagle}
can take over. Within the wait, Figure~\ref{fig:acceptance}c shows that progressively
refreshed KV can support better drafts. During online
drafting, later updates cannot revise drafts already
generated. Our refresh controls show no measurable
acceptance gain on L5 or L10, while in-place updates
avoid the frozen L5 control's extra 1.1\,GiB of KV space
(Appendix~\ref{app:kv-refresh}).
Richer policies could exploit the improving state
through delayed drafting or draft trees~\cite{medusa, eagle2}.
Delayed drafting trades a shorter window for better KV.

\sys uses spare compute on the D-instance, where ordinary decoding is bandwidth-bound.
Prefilling the whole suffix on the D-instance offers another way to use this compute while avoiding
the remote wait~\cite{dynamocondagg, ppd, ampd}.
A natural placement policy combines the two approaches.
When a suffix is short enough to prefill without
substantially delaying ordinary decoding, the D-instance handles it
locally without speculation. Longer suffixes go to the P-instance,
while \sys drafts on the D-instance during the wait.

\sys requires session affinity, the previous request's KV still in cache, and spare decode capacity. Otherwise, the request
takes the non-speculating path. Its gain does not rely on code predictability: 96\% of the draft positions in
Figure~\ref{fig:acceptance}b lie in the agent's natural-language reasoning, before its command. Drafts fail where the
model hesitates (\S\ref{sec:acceptance}), so its gain depends on the model's confidence in its output.

\section{Related Work}
\label{sec:related}

Prefix caching~\cite{vllm, sglang, cachedattention, mooncake} reuses KV only up to the first edit.
Approximate reuse goes beyond it through prompt modules~\cite{promptcache}, cached chunks with selective recomputation~\cite{cacheblend, epic}, or positional re-embedding after truncation~\cite{cachedattention}.
Closest to our setting, Leyline re-rotates the suffix after an agent's span edit~\cite{leyline}, while PatchKV and budgeted repair first recompute edit-local positions~\cite{patchkv, contiguity}.
These methods decode on approximate KV, so outputs can change. \sys also re-positions \invkv, but uses it only to propose tokens and commits only what exact KV confirms.
Cache-aware harnesses preserve reuse by postponing or batching edits~\cite{tokenpilot, cliffcompaction}.
\sys complements them by
lowering the cost of each edit.

Disaggregated serving~\cite{distserve, splitwise, mooncake} and chunked
prefill~\cite{sarathi} protect decodes from long prefills. Agent-aware serving retains KV across
tool calls~\cite{infercept, continuum, kvflow}. An edited request still
waits for exact KV. Mooncake and Splitwise shorten this wait through
layer-wise streaming. \sys turns it into drafting, and the retention
policies above can help keep its draft state resident.

Most speculative drafters need exact prompt KV or their own prefill
before drafting~\cite{leviathan, medusa, eagle}.
Other methods use the target model to draft on sparse, compressed or quantized KV to reduce decoding
traffic~\cite{triforce, magicdec,
quantspec, sparsespec}, or overlap drafting with loading or
transfer~\cite{vericache, lynx}.
\sys drafts on KV already cached for a previous context,
rather than a lossy copy of the current one.

StaleSpec~\cite{stalespec} is concurrent work that shares the same idea of leveraging (re-aligned) \invkv while verifying with the exact KV.
\sys goes beyond this work in several different ways.
First, we observe that streaming the KV to the D-instance as prefill takes place is helpful from both a latency and draft acceptance perspective.
Second, we focus on when to schedule and yield speculations based on varying load to support deployment in real-world P/D disaggregated systems, including identifying draft hesitation as a strong indicator for yielding.

\section{Conclusion}
\label{sec:conclusion}

In this paper, we addressed the problem of reducing the latency penalty observed in existing systems when agents choose to edit contexts such that traditional prefix caching is no longer sufficient.
\sys leverages \invkv state to speculate on a draft, which it verifies after the exact KV arrives.
We identify multiple opportunities to improve this speculation, including streaming of KV updates in-place as drafting continues.
Additionally, \sys addresses how to effectively schedule and yield speculations as session (and request) load varies over time, including using the cumulative acceptance probability to identify when later draft tokens are unlikely to verify.
Our experiments, both on a small testbed and cluster-level simulation, demonstrate the effectiveness of \sys at reducing the latency penalty.

\section*{Acknowledgments}
We used AI assistants and coding agents (ChatGPT, Codex and Claude Code) to help implement \sys and its simulator, run experiments, search related work, and polish and edit parts of the text.

\bibliography{confs_long,refs}
\bibliographystyle{abbrvnat}

\appendix
\renewcommand{\topfraction}{0.85}
\renewcommand{\bottomfraction}{0.6}
\renewcommand{\textfraction}{0.1}
\renewcommand{\floatpagefraction}{0.75}
\renewcommand{\dbltopfraction}{0.85}
\renewcommand{\dblfloatpagefraction}{0.75}
\setcounter{topnumber}{3}
\setcounter{bottomnumber}{2}
\setcounter{totalnumber}{4}
\setcounter{dbltopnumber}{2}

\section{Workloads and Replay}
\label{app:workload}

\subsection{Recording and Selection}
\label{app:recording}

We record SWE-agent~1.1 with its bundled thought--action configuration, which its maintainers describe as similar to that of the SWE-agent paper, keeping the last five observations in full (L5) or the last ten (L10).
Each model runs at temperature~0 on the same 100 SWE-bench Verified issues, drawn with a fixed seed.
A session ends once it exceeds 100 requests, a cap of ours; SWE-agent's paper limits the cost per instance instead.
Twenty issues are held out to fit the agreement tables of the early stop (Appendix~\ref{app:policy-testbed}), and only sessions on the other 80 are replayed.
Requests beyond the context length or whose reply hit the 4{,}096-token cap are left out, as are sessions that end after their first request.
With \mainmodel on \mainwl, this leaves 95 sessions, 77 of them on the replayed issues (2{,}224 requests); the other evaluation sets hold 74 sessions (Llama L10) and 80 each (Gemma and Mistral).

\subsection{Replay Protocol}
\label{app:replay}

Each run replays one workload at a fixed session arrival rate $\lambda$. Sessions arrive open-loop on a seeded
schedule, and within a session each request waits for the previous reply and the recorded tool time. Prompts are the
recorded token sequences, and each reply is generated to its recorded length; drafting does not see this length, so it never cuts a draft window short. A run first fills the system with unmeasured sessions for 400\,s. About $600\lambda$
measured sessions (at least 12) then arrive and run to completion while unmeasured ones keep arriving, and the run ends
at most 900\,s after the last measured arrival. The measured sessions are a prefix of an order that we
rerandomize~\cite{rerandomization} so that its prefixes match the workload within 5\% on six request statistics, among them
the re-prefill per miss request, the reply length and the share of miss requests; all systems replay the same schedule and
orders. A cell of Table~\ref{tab:main} has 12--33 measured sessions and 274--1{,}095 measured requests. We report the ratio
of mean latencies over the requests that both systems complete, paired by request, with 95\% confidence intervals from
1{,}000 bootstrap resamples of sessions. The main configuration at moderate load was run twice, and we pool the
two runs.

Because prompts are the recorded ones, each contains the recorded reply rather than the reply that the D-instance generated, which leaves
\sys somewhat less reusable \invkv than live serving would.

\subsection{Sessions at the Request Limit}
\label{app:loops}

Nine of the 95 sessions reach the request limit. They hold 35\% of the requests and 42\% of the misses, and three of them
loop, repeating a few actions until the limit. We keep all of them: they are the model's behavior under a published
configuration, and removing sessions after the fact by how loop-like they look would select the sample. Without the
nine, a miss re-prefills 3.0k tokens at the median and 8.3k at the 90th percentile, against 3.2k and 11.2k over all 95.
The looping sessions affect only the offline measurements (Table~\ref{tab:long-sessions}): none is among the measured
sessions of \mainwl's cells in Table~\ref{tab:main} or of the single-request ablations, and without them the whole window
of Figure~\ref{fig:acceptance}a is accepted in 63.4\% of miss requests instead of 65\%. The online results thus gain nothing from these sessions, and the offline ones gain little.

\begin{table}[!htb]
\centering
\caption{Offline acceptance on \invkv by session group (the 1{,}400 miss requests of Figure~\ref{fig:acceptance}b, \fixedtie,
32-token window). Figure~\ref{fig:acceptance}a's 65\% whole-window share comes from a different sample of 240 requests
(Appendix~\ref{app:offline}).}
\label{tab:long-sessions}
\vskip 0.1in
\scriptsize\setlength{\tabcolsep}{4pt}
\begin{tabular}{@{}lrrr@{}}
\toprule
Sessions & Requests & Accepted (of 32) & Whole window \\
\midrule
3 looping at the limit & 183 & 30.2 & 81\% \\
6 others at the limit & 392 & 25.5 & 64\% \\
86 below the limit & 825 & 24.2 & 58\% \\
\midrule
All & 1{,}400 & 25.4 & 62.5\% \\
All but the 3 looping & 1{,}217 & 24.6 & 59.7\% \\
\bottomrule
\end{tabular}

\end{table}

\section{Invalidated KV: Additional Measurements}
\label{app:invkv}

\begin{figure*}[t]
\centering
\includegraphics[width=\textwidth]{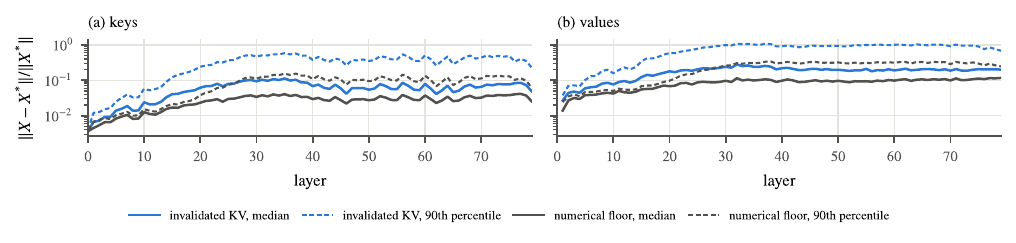}
\caption{Per-token relative KV difference on the 240 miss requests of Figure~\ref{fig:acceptance}: median and 90th
percentile for keys (a) and values (b), with the numerical reference (``floor'') from the same prompt computed at
shifted positions. Layer~0's values are identical and omitted from the logarithmic plot.}
\label{fig:per-layer}
\end{figure*}

\begin{figure*}[t]
\centering
\includegraphics[width=\textwidth]{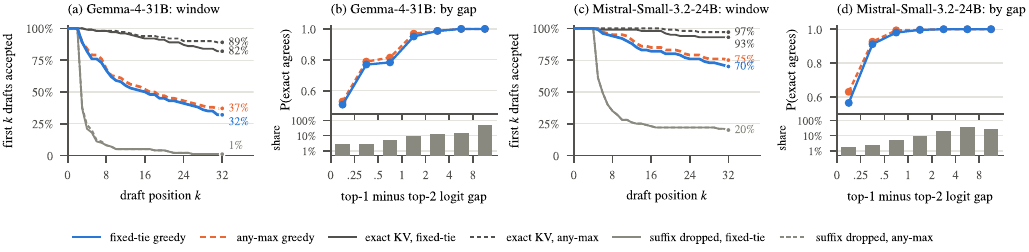}
\caption{Figure~\ref{fig:acceptance}a and~b for Gemma-4-31B (a, b) and Mistral-Small-3.2-24B (c, d): 100 miss requests of
each model's own \mainwl recording, 32-token window, offline. In (b) and (d), the two smallest logit-gap bins hold
46--68 positions each.}
\label{fig:accept-models}
\end{figure*}

\subsection{Offline Study}
\label{app:offline}

The offline measurements of \S\ref{sec:acceptance} replay miss requests of the \mainwl recordings outside the serving system.
For each request, the previous request's KV is computed, aligned with the new prompt as the online system plans it and
re-rotated, and the model drafts greedily for 32 positions; each draft is compared with the exact model's argmax.
Figure~\ref{fig:acceptance}a and~c use 240 requests and Figure~\ref{fig:acceptance}b 1{,}400, both from one seeded random
order of the recording's miss requests; the teacher-forced agreement comes from the 400 of these requests that record the full
next-token distributions. The n-gram drafts use prompt lookup with the longest matching suffix, from four tokens down to
two~\cite{pld}.
A draft position counts as the agent's reasoning (\S\ref{sec:discussion}) if it precedes the reply's first code fence, where SWE-agent's command begins; 96\% of Figure~\ref{fig:acceptance}b's draft positions do.

\subsection{Per-Layer Difference}
\label{app:per-layer}

An edit reaches a retained token's KV only through attention (\S\ref{sec:acceptance}), so re-rotated \invkv is exact in
layer~0 up to rounding and departs from the exact KV only as deeper layers attend to the edited context.
Figure~\ref{fig:per-layer} measures the relative $L_2$ difference $\| X - X^{*} \| / \| X^{*} \|$ at each retained
position and layer of the 240 miss requests of Figure~\ref{fig:acceptance} (240{,}461 positions), against a numerical
reference: the same prompt computed at shifted positions, with its keys re-rotated back.

In layer~0, re-rotated \invkv matches the numerical reference. The median difference first exceeds twice the
reference at layer~14 for keys and layer~12 for values; in the deep layers, it stays at about twice the reference, and
the 90th percentile at three to four times. Re-rotated \invkv thus stays close to the exact KV in every layer, it's a close substitute to draft from.

\subsection{Acceptance on Other Models}
\label{app:accept-models}

Figure~\ref{fig:accept-models} repeats Figure~\ref{fig:acceptance}a and~b for the other two models on 100 miss requests
of each model's own \mainwl recording. Mistral-Small-3.2-24B accepts its whole window on \invkv in 70\% of miss requests
(75\% under \anymax verification), with 27.0 drafts accepted on average against 30.9 on the exact KV. Gemma-4-31B
accepts its whole window in only 32\% (37\%), with 17.2 drafts against 29.1 on the exact KV. Gemma's gap comes mainly
from its chat template, which opens each reply with four thought tokens but renders earlier replies without them, so
the \invkv of the previous reply was computed after tokens that the next prompt does not contain; drafting on KV
computed for the reply as the next prompt renders it accepts 23.0 drafts instead of 17.2. Online at moderate load, 67\%
of Gemma's drafted tokens are accepted (80\% on \mainmodel), and its longer replies keep more requests on the D-instance, so the
staging budget lets only 62\% of its miss requests speculate (86\% on the main workload). Together, these explain why Gemma
gains least in Table~\ref{tab:main}.

\subsection{Refreshing KV While Drafting}
\label{app:kv-refresh}

Figure~\ref{fig:acceptance}c measures drafting from fixed mixtures of invalidated and exact KV. To test updates during
drafting, we compare \sys with a control that drafts from a frozen copy while exact KV arrives, in the single-request
setup of \S\ref{sec:ablation}. On 392 L5 miss requests, \sys accepts 22.6 drafts per request and the control 22.5, both at a
1.155$\times$ miss-request speedup. On 999 L10 miss requests with a cap of 256 drafts, refreshing adds 0.39 accepted drafts per
request (95\% CI $-0.83$ to $1.63$). The gain from refreshing is thus not significant and within the noise of these measurements,
while in-place updates avoid the frozen copy, which needs about 1.1\,GiB more KV space. We still report the offline gain of Figure~\ref{fig:acceptance}c: it guided the design, and drafting policies that
use the improving state could realize it online (\S\ref{sec:discussion}).

\section{Implementation Details}
\label{app:impl}

\subsection{Engine Integration}
\label{app:impl-integration}

\sys is a Python package loaded through vLLM's scheduler and KV-connector extension points. Its decode-side scheduler subclasses vLLM's asynchronous scheduler and applies the policy of \S\ref{sec:policy} before vLLM's batch scheduling, and the connectors on both instances subclass vLLM's NIXL push connector.
A proxy submits each request to the P-instance and the D-instance concurrently and forwards the session identifier that the client attaches, which the D-instance uses to find the previous request of the same session; a request without one never speculates.
Runtime wrappers let the runner's sampler score verification chunks and keep unverified KV out of the prefix cache.

\subsection{Staging, Remap and Verification}
\label{app:impl-staging}

The remap copies the KV of retained positions from the previous request's cached blocks into the request's blocks at their new positions.
vLLM's attention backends store each layer's cache as blocks of rows, one row per position and KV
head holding its key and value.
A fused Triton kernel moves these rows. Since rotary embeddings compose, moving a
key from position $p$ to $p'$ needs only the extra rotation by $\Delta = p' - p$. Values are copied unchanged, and rows
before the edit ($\Delta = 0$) come out bit for bit.
The kernel runs on its own CUDA stream; only after it has finished
does the D-instance release the P-instance's registration and the temporary references to the source blocks. All three evaluated models keep bf16 KV caches, which the kernel requires; the FP8 checkpoint of \mainmodel quantizes weights and activations only.
Rotary variants that change only the frequencies, such as linear scaling, would need only their inverse frequencies.
Layer~0 of the re-rotated \invkv (Appendix~\ref{app:per-layer}) and the KV that the D-instance computes
for an edited span directly after the exact prefix are exact up to rounding, but \sys treats both as
provisional, since the P-instance recomputes them anyway as part of the suffix that verification needs.

After the remap, the D-instance computes the positions with no counterpart in the previous request without new execution code: for
each span, \sys sets the request's bookkeeping so that vLLM sees the tokens before the span as computed and the
prompt as ending at the span's end, and vLLM then runs the span as an ordinary chunked prefill in its batch. Spans run
in position order, the newly appended turn last, whose final step yields the first draft.

When the exact KV has landed, \sys truncates the request to its prompt and its $k$ drafts, sets its computed-token count to $n-1$ for an $n$-token prompt, and re-adds it, so that vLLM schedules the last prompt token and the drafts as one $(k{+}1)$-token chunk in the ordinary batch, with the request's own block tables.
The draft cap keeps this chunk within one $C_s$-token chunk, and a wrapper around the runner's sampler scores its rows.
Under \fixedtie verification, a draft is accepted if it equals its row's argmax and every earlier draft was accepted; under \anymax verification, if its logit equals its row's maximum.
For speculative sampling, the wrapper applies the standard rule~\cite{leviathan}: it accepts
draft $x_i$ with probability $\min(1, p(x_i)/q(x_i))$, where $q$ is the draft distribution recorded while drafting and $p$ the target's, both after vLLM's temperature, top-$k$ and top-$p$ processing, and it draws a rejection's correction from the
normalized residual $\max(0, p-q)$.
Speculative sampling covers requests with a temperature above zero and a top-$k$ of at
most 64.
On a rejection, \sys rewrites the outputs to the accepted drafts and the correction and
recomputes the block hashes from the first rewritten position, so that no rejected draft reaches the prefix cache.
Staging and verification thus run on vLLM's own chunked-prefill path: \sys only edits each request's token bookkeeping and re-queues it, and vLLM's scheduler and kernels do the work unchanged.

\subsection{Layer-wise Streaming}
\label{app:impl-streaming}

The P-instance's connector records, for each prefill step, the blocks that each request's chunk completes. After each attention
layer, it queues the layer's completed blocks for the writer thread of \S\ref{sec:streaming}, which writes them with
vLLM's own descriptors and no notification. When the request finishes, the connector issues vLLM's own final push,
restricted to the blocks that no layer has written (normally only the prompt's last, partly filled block) and only after
every layer write has completed, so that its completion notification still means that all of the request's KV is in
place. If the D-instance's registration reaches the P-instance only after the prefill has finished, the P-instance pushes the request in one piece, as vLLM
does. A sliding-window group streams only the blocks of its last window.

The implementation supports, and requires, tensor parallelism of one and the same block size on both instances, the
configuration of our experiments; other requests are pushed in one piece, as in vLLM. Neither is a requirement of the
design. Under tensor parallelism, each prefill rank would stream its heads' layers to the decode rank that holds the
same heads, as vLLM's push already pairs ranks, and re-rotation acts on each head separately; different block sizes
change only the addresses of each layer's write.

\subsection{Scratch Block Management}
\label{app:impl-scratch}

Scratch blocks isolate the D-instance's prompt-row writes from the P-instance's transfers. For each KV cache group, the D-instance substitutes scratch blocks
in its staging and drafting block table for every block that intersects a prompt span it computes, including the last
prompt block, whose unused rows hold the first drafts; the P-instance keeps writing the original registered blocks. Scratch blocks
come from the ordinary KV pool, are never published to the prefix cache and follow the limits of
Appendix~\ref{app:policy-testbed}. At verification, the request re-enters the batch with its original block tables, so
nothing is copied back, and its scratch blocks are freed once no outstanding D-instance step uses them.

\subsection{Hybrid Attention Models}
\label{app:impl-hybrid}

Gemma-4-31B interleaves sliding-window layers, which attend only to the previous 1{,}024 positions, with
full-attention layers, five to one. vLLM's hybrid KV cache manager keeps them in separate KV cache groups and frees a
sliding-window layer's blocks once they fall out of every running request's window. \sys plans, remaps and isolates each group on its own.

The plan starts from the positions the D-instance must compute. Walking back from the tail, an edited span within one window of a
span that the D-instance already computes is \emph{near}: sliding-window layers read it while the D-instance computes, so the D-instance computes it too. Any
other edited span is \emph{far}: no sliding-window layer reads it during staging or drafting, so the D-instance skips it, and the
full-attention layers read the previous request's KV at its positions unchanged ($\Delta = 0$). A request whose far span lies
past the previous request's rows does not speculate. The full-attention group then needs every retained row, while a
sliding-window group needs only the window of rows before each span that the D-instance computes.

vLLM keeps a finished request's
sliding-window KV cached only at checkpoints; the D-instance runs with a checkpoint every 32 tokens in every system of our Gemma
experiments, so that the previous request's last window stays cached. If no more than 32 rows at the far edge of a window
are no longer cached, \sys fills them with zeros, which affects only the drafts; otherwise the request does not speculate.
While a request speculates, \sys caps sliding-window eviction at the prompt's last token, so that blocks the P-instance may still
write or the D-instance may read stay allocated.

These rules cost Gemma little. At moderate load, 11 of the 639 miss requests that the D-instance considered for speculation (1.7\%)
did not speculate because a window row they needed had already been freed. Its staging finished in 0.16\,s at the median, against 0.29\,s
on \mainmodel in the main-table runs, since its speculations staged 40\% fewer tokens at the median.

\subsection{Asynchronous Execution and Safe Fallback}
\label{app:impl-async}

Under asynchronous scheduling, the next step is already scheduled when a step's output arrives. \sys therefore records
the role of each managed request's work in every step (a span, a draft or verification) and interprets outputs by that
record, so a decision takes effect at the next step that can still be scheduled. The D-instance's connector reports a completed transfer at the end of the step that received
it, and only if that step reports no KV-load error; a request with drafts then enters verification one step before
vLLM's own update would promote it.

A speculation whose registration is still withheld 10\,s
after admission returns to the ordinary path, well within the 30\,s for which the P-instance keeps a finished prefill's KV.

When vLLM preempts a speculating request under memory pressure, \sys drops the drafts and returns the request to vLLM's
remote-KV wait path, retaining the registered blocks until the P-instance's writes finish. Abandoned speculations use the same
fallback. If the request is aborted while the P-instance may still write, its registered blocks remain allocated until the transfer
completes. Scratch blocks are reclaimed after their D-side users finish.

\section{Decode-Side Policy}
\label{app:policy}

\subsection{Rules and Defaults}
\label{app:policy-testbed}

A miss request speculates only if it passes every admission check when it reaches the D-instance; otherwise it takes the ordinary path
at once and never queues for speculation. It needs a free decode batch slot that no ready request is waiting for, and
every block of the previous request that staging needs must still be cached (Appendix~\ref{app:impl-hybrid} covers sliding
windows). At least 256 tokens of \invkv must be reusable beyond the exact prefix that the D-instance has cached. The $s$ tokens that
the D-instance must compute during staging and the number $B$ of running requests, the request included, must satisfy $sB \le K$, with
$K=4{,}096$. Scratch blocks in use must stay within 10\% of the D-instance's KV pool and leave at least 5\% of the pool free, and the
verifier must support the request's decoding parameters (Appendix~\ref{app:impl-staging}).
A request whose exact KV arrives before its first draft takes the ordinary path.

A drafting request adds one decode row per step to the ordinary batch. Drafting ends when the exact KV arrives, at a
stop token, at the draft cap, or by the early stop. The draft cap keeps the last prompt token and all drafts within one
verification chunk of $C_s$ tokens, so a request drafts at most $C_s - 1$ tokens (111 at the main configuration). For
greedy decoding, the early stop maps each draft's top-1 minus top-2 logit gap to an agreement probability $g$,
multiplies these into the cumulative acceptance probability, and stops drafting when it falls below $\theta = 0.3$.
Speculative sampling drafts until the exact KV arrives, a stop token or the cap.

Table~\ref{tab:agreement} lists each model's agreement table, which maps seven logit-gap bins to agreement
probabilities. We fit each table offline on 150 miss requests from the 20 held-out issues (Appendix~\ref{app:recording}).
Each request drafts greedily for 32 positions on its \invkv, and each draft is compared with the exact model's argmax. A
bin's value is its agreement rate over the positions reached without an earlier mismatch, so each $g$ is conditional on
all earlier drafts having been accepted, and their product estimates the cumulative acceptance probability.

\begin{table}[!htb]
\centering
\caption{Agreement tables of the early stop: agreement probability $g$ by the draft's top-1 minus top-2 logit gap,
fitted per model on the held-out issues.}
\label{tab:agreement}
\vskip 0.1in
\footnotesize\setlength{\tabcolsep}{3.5pt}
\begin{tabular}{@{}lrrrrrrr@{}}
\toprule
Logit gap & $<$.25 & .25--.5 & .5--1 & 1--2 & 2--4 & 4--8 & $\ge$8 \\
\midrule
Llama-3.3-70B & 0.63 & 0.69 & 0.89 & 0.97 & 1.00 & 1.00 & 1.00 \\
Mistral-Small-3.2 & 0.70 & 0.92 & 0.99 & 1.00 & 1.00 & 1.00 & 1.00 \\
Gemma-4-31B & 0.57 & 0.78 & 0.86 & 0.95 & 1.00 & 1.00 & 1.00 \\
\bottomrule
\end{tabular}
\end{table}

When ready requests need more slots than are free, \sys first releases the slots held by requests that have stopped
drafting and are waiting for their exact KV; if more are needed, drafting requests yield in increasing order of their
cumulative acceptance probability. A request yields only after its staging has finished, and it does not resume
drafting. It keeps its blocks and drafts, and once its exact KV has landed, its verification is scheduled like an
ordinary request whose KV has arrived.

The defaults follow from the D-instance's step structure. $C_s$ is 128 minus the D-instance's batch limit (112 at the main configuration's 16),
so that one staging chunk and a decode row from every other request fit in one 128-row tile of the FP8 matrix
multiplication; above 128 tokens, a step switches to a slower configuration (the $G(n)$ term of
Equation~\ref{eq:dstep}). Against unchunked staging at moderate load, these chunks start drafting later (308 against
233\,ms at the median) but avoid most steps above 128 tokens (103 against 528), and unchunked staging raises miss-request
latency by 1.1\% (95\% CI 0.2--3.2\%). The 256-token minimum excludes requests whose reusable context or re-prefill is too
short to repay the fixed cost of staging and verification. The budget $K=4{,}096$ lets a request stage up to 4{,}096 tokens
on an idle D-instance and 256 beside 15 other requests, close to the break-even budget of Appendix~\ref{app:policy-budget}. The
scratch limits exist because vLLM cannot reclaim scratch blocks by preempting their requests, and $\theta = 0.3$
balances discarded against accepted drafts (Appendix~\ref{app:policy-sensitivity}). Every configuration of
Table~\ref{tab:main} uses these defaults; only the agreement table differs by model.

Admission is immediate: a queue for speculation would mostly admit work after waiting for capacity on a busy D-instance,
exactly where the policy aims to speculate less, while the P-instance keeps shortening the window left to repay staging. The checks
respond only to the D-instance's slots and staging burden, not to the arrival rate or the P-instance's load, since a busy P-instance can still leave the D-instance
with spare capacity and a long wait to fill.

Table~\ref{tab:admission} counts the admission decisions on the miss requests of the main configuration. At a batch limit
of 16, only the budget, the reuse minimum and eviction of the previous request's KV refuse requests up to capacity; no request
is refused for want of a slot or scratch space. The reuse minimum excludes 0.2\% of the miss requests, and the budget
refuses more as the D-instance's batch fills, 14\% at moderate load and 31\% at full load. The slot rules act only when the D-instance is tight: at
a batch limit of 2 and its capacity of 0.02 sessions/s, 155 miss requests are refused for want of a slot, against 342
that speculate.

\begin{table}[!htb]
\centering
\caption{Admission of miss requests on the main configuration (D-instance batch limit 16) by load, over every miss request of the run
that reaches admission, measured or not; past capacity: 0.045 sessions/s; moderate load pools two arrival schedules.
Spec.: miss requests that speculate; Evicted: a block of the previous request that staging needs was no longer cached; Cut:
share of speculations cut short within three drafts because the exact KV arrived.}
\label{tab:admission}
\vskip 0.1in
\footnotesize\setlength{\tabcolsep}{2.5pt}
\begin{tabular}{@{}lrrrrrrr@{}}
\toprule
 & & & \multicolumn{4}{c}{Refused} & \\
\cmidrule(lr){4-7}
Load & Misses & Spec. & Budget & Evicted & Scratch & Slot & Cut \\
\midrule
Light & 280 & 274 & 6 & 0 & 0 & 0 & 1.8\% \\
Moderate & 1{,}247 & 1{,}070 & 173 & 4 & 0 & 0 & 4.3\% \\
Full & 1{,}103 & 755 & 344 & 4 & 0 & 0 & 4.0\% \\
Past cap. & 1{,}271 & 652 & 402 & 54 & 163 & 0 & 3.5\% \\
\bottomrule
\end{tabular}

\end{table}

\subsection{Deriving the Staging Budget}
\label{app:policy-budget}

The rule $sB \le K$ follows from the D-instance's step-time model (Equation~\ref{eq:dstep}). Let a miss request $c$ arrive while $B-1$
other requests run, stage $s$ tokens, draft $k$ tokens and have $A$ of them accepted. Assume that the other requests
stay in the batch, that $c$'s exact KV lands at the same time with or without speculation, and that steps carrying
staging stay within 128 tokens. Staging then lengthens the steps that carry it by $a_p s$ in total, each discarded draft
adds a row that costs $\delta = a_r + a_c\ell_c$ (with $\ell_c$ the context of $c$ in thousands of tokens), and each
accepted draft spares $c$ one decode step of $\tau$ after its exact KV lands. Admitting $c$ lowers the sum of the
latencies of $c$ and the other requests when
\begin{equation}
(B-1)\bigl[a_p s + (k-A)\,\delta\bigr] \le A\tau .
\label{eq:breakeven}
\end{equation}
The right side does not depend on $B$, so the staging that a request can afford falls in inverse proportion to the batch.
Neglecting the discarded drafts, which the early stop limits, and counting $c$ in $B$ gives $sB \le K^{*}_c = A\tau/a_p$,
the form of \sys's rule.

On the main configuration at moderate load, a speculating miss request drafts 26.7 tokens and accepts 21.0, \base decodes
these requests at $\tau = 26.3$\,ms per output token (the requests of Figure~\ref{fig:micro}), and $a_p = 0.148$\,ms per token
(Appendix~\ref{app:sim-validation}), so $K^{*}_c \approx 3{,}700$. Keeping the discarded drafts, with $\delta =
1.8$--$2.5$\,ms, the measured lengthening of steps that carry only drafting rows, lowers it to 2.3--3.5k for $B$ between 4 and 16. $K^{*}_c$ cannot be
computed at admission, so \sys uses the constant $K = 4{,}096$, slightly above these values, which errs toward admitting
requests. An error in $K$ costs little: in the runs of Table~\ref{tab:policy-sweep}, doubling $K$ changes the all-request
speedup by at most 0.004 and halving it by at most 0.018.

\subsection{Parameter Sensitivity}
\label{app:policy-sensitivity}

We vary the stopping threshold $\theta$ and the staging budget $K$ separately in the simulated 1:1 cluster at moderate
and 90\% load, with three arrival schedules per setting and every other control fixed (Table~\ref{tab:policy-sweep}).
The no-budget control removes only the $sB \le K$ check; it keeps slot availability, KV availability and the other
admission checks.

\begin{table}[!htb]
\centering
\caption{Parameter sweeps in the simulated 1:1 cluster (96 GPUs, three arrival schedules per setting). Values are
latency speedups over \base. Others denotes non-speculating requests.}
\label{tab:policy-sweep}
\vskip 0.1in
\small\setlength{\tabcolsep}{3pt}
\begin{tabular}{@{}lrrrr@{}}
\toprule
 & \multicolumn{2}{c}{Moderate load} & \multicolumn{2}{c}{90\% load} \\
\cmidrule(lr){2-3}\cmidrule(lr){4-5}
Setting & Miss & Others & Miss & Others \\
\midrule
$\theta=0.1$ & 1.162 & 1.024 & 1.193 & 1.071 \\
$\theta=0.3$ & 1.161 & 1.026 & 1.194 & 1.076 \\
$\theta=0.5$ & 1.155 & 1.024 & 1.195 & 1.083 \\
No early stop & 1.164 & 1.027 & 1.194 & 1.074 \\
\midrule
$K=2{,}048$ & 1.134 & 1.048 & 1.182 & 1.091 \\
$K=4{,}096$ & 1.161 & 1.026 & 1.194 & 1.076 \\
$K=8{,}192$ & 1.169 & 1.008 & 1.198 & 1.037 \\
No budget & 1.166 & 1.005 & 1.189 & 1.000 \\
\bottomrule
\end{tabular}
\end{table}

The budget trades miss-request gain against the latency of requests that do not speculate. At 90\% load, raising $K$ from
2{,}048 to 8{,}192 lifts miss-request speedup from 1.182$\times$ to 1.198$\times$ but lowers that of non-speculating requests
from 1.091$\times$ to 1.037$\times$. Removing the budget leaves these requests no gain, and at 2:1 and 3:1 makes them
slower than under \base at moderate load (0.88--0.92$\times$). The default sits between these extremes.

The threshold changes how much draft work is discarded, not the latency. Across $\theta \in \{0.1, 0.3, 0.5\}$ and
without early stopping, miss-request speedup varies by less than 1\% at either load, and non-speculating requests are almost
unaffected. Against disabling the early stop, the default discards 23--35\% fewer drafts in each P:D allocation and
load, 30\% pooled, while keeping 97\% of the accepted drafts (Table~\ref{tab:policy-draft-work}). A lower threshold,
$\theta = 0.1$, rarely stops drafting and cuts discarded drafts by only 3\%; a higher one, $\theta = 0.5$, cuts them by
58\% but keeps only 90\% of the accepted drafts.

\begin{table}[!htb]
\centering
\caption{Confidence early stopping in 96-GPU clusters, against disabling it. Top: the default $\theta=0.3$ per
allocation and load; accepted/spec.\ gives mean accepted drafts per speculation (default / disabled). Bottom: pooled
over the eight settings.}
\label{tab:policy-draft-work}
\vskip 0.1in
\small\setlength{\tabcolsep}{3pt}
\begin{tabular}{@{}llrrr@{}}
\toprule
P:D & Load & Discarded & Total accepted & Accepted/spec. \\
 & & reduced by & retained & default / disabled \\
\midrule
1:2 & Moderate & 35.4\% & 98.1\% & 22.3 / 22.5 \\
 & 90\% & 34.1\% & 96.7\% & 26.0 / 26.7 \\
1:1 & Moderate & 30.3\% & 99.1\% & 19.7 / 19.8 \\
 & 90\% & 28.5\% & 96.2\% & 22.7 / 23.2 \\
2:1 & Moderate & 29.3\% & 94.5\% & 16.0 / 16.3 \\
 & 90\% & 24.1\% & 98.0\% & 18.2 / 18.5 \\
3:1 & Moderate & 23.3\% & 98.8\% & 13.8 / 13.8 \\
 & 90\% & 22.8\% & 98.0\% & 13.1 / 13.2 \\
\midrule
\multicolumn{2}{@{}l}{Pooled, $\theta=0.1$} & 2.6\% & 99.9\% & \\
\multicolumn{2}{@{}l}{Pooled, $\theta=0.3$} & 30.1\% & 97.1\% & \\
\multicolumn{2}{@{}l}{Pooled, $\theta=0.5$} & 57.9\% & 89.9\% & \\
\bottomrule
\end{tabular}
\end{table}

\subsection{Alternative Policies}
\label{app:policy-alternatives}

Admission, early stopping and yielding are separate decisions from constructing \provkv and verifying drafts, and
more informed rules could use the same KV and verification mechanisms. Our default uses locally available signals and
fixed thresholds to keep deployment simple. These estimates govern how much to speculate, while exact verification
still determines which tokens are committed.

Early stopping also frees decode compute for richer drafting. Each draft occupies a row in a step of the D-instance's batch, so a
per-request budget of draft rows bounds the compute that speculation takes from co-running requests. Under such a
budget, the rows that the early stop saves, 30\% of the discarded drafts at the default
(Appendix~\ref{app:policy-sensitivity}), could fund delayed drafting, which waits for more exact KV before drafting, or
draft trees, which branch where the model hesitates (\S\ref{sec:discussion}). The headroom for such policies is
bounded: if every draft were accepted, a speculating miss request at moderate load would finish only 0.15\,s sooner
(Appendix~\ref{app:metrics}).

\S\ref{sec:discussion} suggests combining \sys with prefilling short suffixes on the D-instance~\cite{dynamocondagg}. The staging
budget already makes this decision for the part of a request that the D-instance computes while it stages: the D-instance computes $s$ tokens only
while $sB \le K$. A combined policy applies the same test to the whole re-prefill. A request whose re-prefill of $r$
tokens satisfies $rB \le K$ is prefilled entirely on the D-instance, skipping the P-instance's queue and the transfer, while longer re-prefills
go to the P-instance and speculate as now. On the main workload, 59.5\% of miss requests re-prefill at most 4{,}096 tokens and would
stay on an idle D-instance, but only 5.4\% beside three other requests and 0.1\% beside 15. The two approaches cover different requests. Local prefill removes the P-instance from short re-prefills, whereas a long
re-prefill on the D-instance would slow every co-running request for many steps: by the D-instance's cost model (Equation~\ref{eq:dstep}),
re-prefilling the median 3.1k tokens adds about 0.46\,s to the steps that the whole batch waits through. \sys serves
these long re-prefills, keeping their recomputation on the P-instance and drafting during the wait. Appendix~\ref{app:sim-combine}
evaluates this rule and Dynamo's fixed threshold together with \sys in simulated clusters.

\section{Evaluation Setup}
\label{app:eval-setup}

\subsection{Baseline and Transfer Path}
\label{app:baseline}

Our baseline uses vLLM's NIXL connector in push mode, in which the P-instance writes a request's KV into blocks that the D-instance has
registered, rather than the default pull mode, in which the D-instance reads it once the prefill has finished. vLLM's developers
report that push mode is faster under load~\cite{vllmpush}, and \sys is built on it, so the comparison isolates \sys's
changes. All systems share the engine configuration and the proxy: the P-instance schedules up to four requests and 8{,}192 tokens
per step, and the D-instance has a batch limit of 16 unless stated otherwise. Apart from speculation, \sys's transfer path differs
only in the P-instance's connector (Appendix~\ref{app:impl-streaming}), chiefly in that it pushes each layer's KV as soon as it has
been computed and its writer notices the final write's completion at once, where vLLM checks once per P-instance step. At
moderate load, pushing the KV after the prefill and noticing the completion take \base 0.39\,s on average and \sys
under 1\,ms.

\subsection{Measuring the Gain}
\label{app:metrics}

A miss request's penalty under \base is the time from its arrival until its exact KV lands on the D-instance, and its folded share is
the net reduction in request latency divided by this penalty, so it accounts for staging, verification and interference.
Since \base does not record when each request's KV lands under load, Figure~\ref{fig:micro} estimates the landing as the P-instance's
response time plus the P-instance's mean push-and-notice time (Appendix~\ref{app:baseline}), and the P-instance's queueing time as the mean over
all requests. Of the 0.81\,s that a speculating miss
request saves at moderate load, 0.39\,s comes from the transfer path and 0.42\,s from accepted drafts; if every draft were
accepted, the request would finish only 0.15\,s sooner, so \sys realizes about 85\% of this ideal saving. Of the speculating miss requests, 73\% accept their whole window and
3\% accept no draft.

The single-request ablations of \S\ref{sec:ablation} replay the first 18 measured sessions of the main workload one
request at a time (392 miss requests). A \base run that cuts every reply to its first token gives each miss request's wait for
the exact KV, from which we subtract one decode step: the penalty is 1.04\,s on average. Folded share and speedup are
ratios of sums over these requests, with 95\% session-bootstrap confidence intervals: \sys hides 49\% [42\%, 54\%] of the
penalty at a 1.155$\times$ [1.138, 1.181] speedup, and n-gram drafts 18\% [10\%, 27\%].

\subsection{Capacity Estimation}
\label{app:capacity}

Capacity is a common load reference for comparing systems (\S\ref{sec:setup}), estimated from \base's runs alone
for each model, workload and D-instance batch limit. Let $m$ be the median request latency at low load. We split each run's requests,
measured and background, into 100-second windows by completion time. A window fails if every completed request in it
takes more than $10m$, and the highest tested rate before the first failure gives an initial estimate. Since a fast request can let a window pass while requests accumulate, we also check
for unfinished measured sessions, requests left unissued at the deadline, and growth in the P-instance's queue over the arrival
window. These checks lower one main-table estimate, Mistral L10's, from 0.040 to 0.035 sessions/s, since both runs at
0.040 pass the window test but leave 45--52 requests unissued. With D-instance batch limits of 6, 4 and 2, the capacities are at
least 0.04, 0.035 and 0.02 sessions/s, the last two set by these checks. On the main configuration, every measured session finishes at 0.040 sessions/s, while at
0.045 four of 27 do not (Appendix~\ref{app:beyond-capacity}).

The window test is robust to its parameters. On the same baseline runs, failing a window when at least half of its
requests exceed $10m$ (in windows with at least five completions) leaves all six main-table estimates unchanged, and
tightening the bound to $5m$ lowers only Llama L10's and Mistral L10's, by 0.005 sessions/s, the latter to the adopted
0.035. Targets over a whole run, which require 90\% of its completed requests to meet the bound, are less suited. At $10m$
they admit 0.045 sessions/s for Gemma L10 and Mistral L5, where requests are left unissued, and at $5m$ they mostly
reflect the workload's own latency spread, since even at light load only 92\% of Gemma L5's completed requests finish
within $5m$. We therefore pair the window test with the completion and queue checks.

Load levels come from a week of Azure's coding traffic~\cite{dynamollm} (1{,}008 ten-minute windows). Relative to the
99th-percentile window, its 25th and 75th percentiles are 0.120 and 0.518, which give light (12\%) and moderate (52\%)
load as fractions of each configuration's capacity; full load is the capacity itself. On the main configuration, the three loads are 0.005, 0.021 and 0.04
sessions/s.

\subsection{Past Capacity}
\label{app:beyond-capacity}

\begin{figure}[!htb]
\centering
\includegraphics[width=\columnwidth]{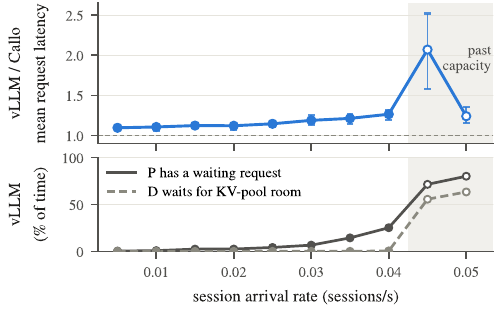}
\caption{The main configuration (D-instance batch limit 16) up to and past its capacity of 0.04 sessions/s. Top: \base/\sys
mean request latency (bars: 95\% confidence intervals). Bottom: share of time under \base that the P-instance has a waiting request and
that the D-instance waits for room in its KV pool. Hollow and shaded: past capacity, where the run left measured requests unissued or
unfinished.}
\label{fig:beyond-capacity}
\end{figure}

Figure~\ref{fig:beyond-capacity} extends Figure~\ref{fig:D_bottleneck} past the main configuration's capacity of 0.04
sessions/s. At 0.045 sessions/s, sessions pile up under \base: by the end of the run, \base has left 90 of the 754
measured requests unissued and 4 of the 27 measured sessions unfinished, against 3 requests and one session under \sys. Under
\base, the P-instance queues 71\% of the time there, and the D-instance does not run full: it waits for room in its KV pool (56\% of the time),
not for a batch slot. At 0.05 sessions/s, both systems are overloaded.

Past capacity, \base's backlog feeds itself through the P-instance's prefix cache (Table~\ref{tab:beyond-capacity}). While the
measured sessions arrive, the two systems keep similar numbers of sessions in flight and the P-instance's prefix-cache hit rate is
71--72\% under both. Afterwards, \base's measured sessions finish later, so more sessions compete for the P-instance's cache (29 in
flight at the peak, against 12 under \sys); the P-instance's hit rate falls to 17\%, against 49\% under \sys, and the P-instance recomputes more.
Since \base's queue keeps growing, the ratio of the two systems' request latencies depends on where the run is cut, so we
report this point as a stability result rather than a gain.

\begin{table}[!htb]
\centering
\caption{The P-instance at 0.045 sessions/s, one step past the main configuration's capacity, against \base at capacity (0.04), one
run each, from the end of the warm-up to the end of the run. Sessions in flight, measured or not: at the end of the
arrival window $\to$ peak afterwards.}
\label{tab:beyond-capacity}
\vskip 0.1in
\small
\begin{tabular}{@{}lrrr@{}}
\toprule
 & \multicolumn{2}{c}{0.045 sessions/s} & 0.04 \\
\cmidrule(lr){2-3}\cmidrule(lr){4-4}
 & \base & \sys & \base \\
\midrule
P-instance cache hit rate & 47.7\% & 60.4\% & 70.6\% \\
P-instance tokens computed & 7.64M & 7.11M & 5.20M \\
Sessions in flight & 13 $\to$ 29 & 11 $\to$ 12 & 11 $\to$ 11 \\
Measured sessions finished & 23 of 27 & 26 of 27 & 24 of 24 \\
Measured requests unissued & 90 & 3 & 0 \\
\bottomrule
\end{tabular}

\end{table}

\section{Cluster Simulation}
\label{app:sim}

\subsection{Simulator}
\label{app:sim-design}

Existing simulators have succeeded in modeling disaggregated serving and prefix caching~\cite{vidur, llmservingsim, frontier}, but they cannot easily support modeling the speculation during a miss or the decode-side policy in our system, and they replay prompts that only grow at their end.
We therefore build our own, with the usual practice in LLM serving simulators~\cite{vidur, distserve, splitwise}: cost models fitted to measurements on the target GPU give the time of each engine step, and an event-driven model of the serving stack's schedulers drives them.
A run is deterministic given its configuration and seeds.

P-instances and D-instances follow vLLM~0.29. Prefill instances run its synchronous scheduler with chunked prefill and hold a request's
blocks until its KV has been sent. Decode instances run its asynchronous scheduler, admit a request whose KV is remote
only while a batch slot and KV room are free, and put such requests and \sys's re-entering requests first in the waiting
queue. Each instance manages its KV as vLLM does, with a pool of 16-token blocks, a prefix cache over them and LRU
eviction of cached blocks. Block hashes are salted per session (the usual default for isolating sessions), so that an
edit in the middle of a prompt misses exactly where it does on the testbed, and a finished request's prompt and reply stay
in the D-instance's prefix cache as evictable blocks. \base sends a request's KV in one write after its prefill and notices the
completion when the P-instance's running step ends; \sys sends one write per layer, and its writer thread notices the completion.
\sys's decode-side policy (admission, staging, drafting, early stop, yield and verification) is ported rule by rule, and
its \invkv mapping, constants and agreement tables are imported from the system unchanged. Speculation admission checks
that the required donor blocks remain cached, scratch use counts against the 10\% cap and is released at verification,
and the D-instance's pool under \sys is smaller by its fixed reserve (\S\ref{sec:memory}). Fewer than 1\% of miss requests find the
previous request's KV gone on arrival, but at 2:1 and 3:1 up to 12\% and 13\% lose it while waiting on the D-instance.

The simulator replays recorded replies and speculation outcomes. A record with $a$ of $k$ drafts accepted gives
$\min(k_{\mathrm{sim}},a)$ accepted drafts when $a<k$; a fully accepted record is extended with longer records of the
same request and then with rejection probabilities estimated from pooled records (Kaplan--Meier). A request with no recorded
speculation borrows a record from one of the five requests closest in invalidated-KV length, and the logit gaps of the
early stop are replayed from the same records.
Step times come from cost models of vLLM~0.29 serving \mainmodel on one H200 per instance.
A decode step schedules $r$ requests and $n$ tokens in total, of which $n_p$ are computed by local prefill and $n_v$ by
verification, and request $j$ reads $\ell_j$ thousand tokens of context. Its time in milliseconds is
\begin{equation}
t_D = a_0 + a_r r + a_c \sum_j \ell_j
      + a_p n_p + a_v n_v + a_t G(n),
\label{eq:dstep}
\end{equation}
where
\[
G(n)=
\begin{cases}
0, & n \le 128,\\
\lceil n/256\rceil, & n>128.
\end{cases}
\]
The linear terms approximate step times over the batch compositions in our experiments, and $G(n)$ models the jump above
128 scheduled tokens, where vLLM's FP8 matrix multiplication switches to 256-row groups, and the group boundaries after
it. A prefill step computes a chunk of $c_i$ tokens for request $i$ after $x_i$ tokens of context, $n=\sum_i c_i$ in
total, and takes
\begin{equation}
\begin{aligned}
t_P = {} & b_0 + b_t G(n) \\
         & {} + 10^{-6}\sum_i
           \left(b_x c_i x_i + b_c c_i^2/2\right)
\end{aligned}
\label{eq:pstep}
\end{equation}
milliseconds, where the two attention terms count token pairs with the preceding context and within the chunk, $b_x$
and $b_c$ are in milliseconds per million token pairs, and $b_0=a_0$.

Each instance has its own 400\,Gb/s RDMA interface~\cite{alibabahpn}, which it divides equally among its active
transfers. A transfer runs at the smaller of its shares at the P-instance and the D-instance, and a non-blocking switch confines contention to
these interfaces. Transfer volume counts the blocks missing on the D-instance, at 327,680 bytes per token for \mainmodel, and HTTP,
API-server, registration and polling overheads of a few milliseconds follow testbed measurements.
Appendix~\ref{app:sim-sensitivity} evaluates 200 and 100\,Gb/s interfaces and the P-instance and the D-instance connected by NVLink on one node.

A router selects the P-instance and the D-instance independently for each request. By default, a cache-aware policy inspired by SGLang's
router~\cite{sglang} returns a request to its session's previous instance of each type if that instance has at most four
more running and waiting requests than the least-loaded one, and otherwise, as for new sessions, sends it to the
least-loaded instance. This favors cache reuse on both P-instances and D-instances; a request proceeds without speculation if its D-instance lacks
reusable \invkv. \base and \sys use the same router, and Appendix~\ref{app:sim-sensitivity} compares it with routing by
KV cost, as in Dynamo, and by load alone.

For each configuration, we estimate \base's capacity with a per-instance version of the testbed's window test
(Appendix~\ref{app:capacity}). At each P-instance and D-instance, a 100-second window fails if it holds at least ten requests and
all exceed ten times \base's low-load median latency, and we take the highest tested rate that passes and at which
\base finishes every measured session on all three arrival schedules. This capacity is the load reference for all
variants, including the 12\% and 52\% points. Weekly speedups (\S\ref{sec:sim}) place the 99th-percentile window of
Azure's week~\cite{dynamollm} at capacity and aggregate the sweep at each window's load, weighted by the window's request
count and \base latency.

\subsection{Calibration and Validation}
\label{app:sim-validation}

We fit the decode model, Equation~(\ref{eq:dstep}), on 491,769 busy steps at a D-instance batch limit of 16. In milliseconds,
with context in thousands of tokens, the fitted coefficients are $a_0=22.45$, $a_r=0.111$, $a_c=0.108$, $a_p=0.148$,
$a_v=0.092$ and $a_t=12.3$. Per-step relative error is 1.2\% at the median and 4.2\% at the 90th percentile. Steps that
carry staging or verification rows, the ones \sys adds, have a median error of 5.4\%, against 1.1\% for decode-only
steps.

The P-instance does not log individual steps, so we fit its model, Equation~(\ref{eq:pstep}), to the prefill times of 615 isolated
low-load requests, minimizing relative error. With $b_0=a_0=22.45$, the fit gives $b_t=30.8$, $b_x=5.16$ and $b_c=13.7$,
with a relative error of 1.9\% at the median and 7.6\% at the 90th percentile. The same requests' transfer times give an
effective NVLink bandwidth of 69.2\,GB/s, which the testbed validation and the NVLink point of
Appendix~\ref{app:sim-sensitivity} use.

\begin{figure*}[t]
\centering
\includegraphics[width=\textwidth]{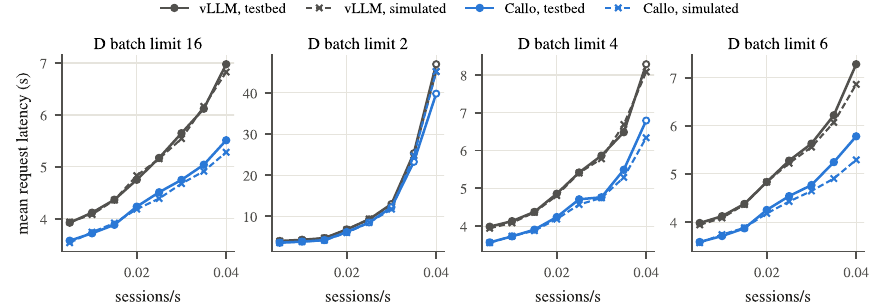}
\caption{Mean request latency against the session arrival rate on the testbed's configuration (one P-instance, one D-instance), measured
and simulated, for \base and \sys at each D-instance batch limit (hollow: testbed points past the configuration's capacity).}
\label{fig:sim-validation}
\end{figure*}

We validate end-to-end behavior by simulating every testbed run (one P-instance, one D-instance) up to 0.04 sessions/s with its own
arguments, 32 load points over D-instance batch limits of 16, 2, 4 and 6, three times each. The runs at batch limits of 2, 4 and
6 are held out from the fit, and each simulated run draws accepted drafts only from the other runs' recorded
speculations. Table~\ref{tab:sim-validation} and Figure~\ref{fig:sim-validation} compare simulation and testbed. Up to
each configuration's capacity, the simulator reproduces \base's per-request latency within 2.2--3.2\% at batch limits of
16, 2 and 4, and within 5.7\% at 6, whose largest error is at the highest rate (0.04 sessions/s). The simulated gain
lies inside the testbed's 95\% confidence interval at 26 of the 32 points, with a median error of at most 0.004 at every
batch limit. Past the main configuration's capacity, at 0.045 and 0.05 sessions/s (Figure~\ref{fig:beyond-capacity}),
the simulator gives \base a mean request latency of 12.4\,s and 20.9\,s, against 13.5\,s and 20.4\,s on the testbed.

\begin{table}[!htb]
\centering
\caption{Simulator against the testbed, by D-instance batch limit. \base latency: largest per-request error over the load points
up to capacity; gain inside the testbed's 95\% CI: out of all load points with a testbed CI; gain error: median over
the points up to capacity; speculating requests and accepted drafts: error in their counts, pooled over the points
up to capacity.}
\label{tab:sim-validation}
\vskip 0.1in
\footnotesize\setlength{\tabcolsep}{4pt}
\begin{tabular}{@{}lrrrr@{}}
\toprule
D-instance batch limit & 16 & 2 & 4 & 6 \\
\midrule
\base latency error & 2.2\% & 2.2\% & 3.2\% & 5.7\% \\
Gain inside testbed CI & 7/8 & 6/8 & 6/8 & 7/8 \\
Gain error & $+0.004$ & $-0.001$ & $+0.003$ & $+0.004$ \\
Speculating requests & $-1.1$\% & $+0.0$\% & $+0.5$\% & $+0.4$\% \\
Accepted drafts & $-3.9$\% & $-4.0$\% & $-1.6$\% & $-4.4$\% \\
\bottomrule
\end{tabular}

\end{table}

\subsection{Speculation and the Transfer Path under Load}
\label{app:sim-stream}

Comparing \sys with \sysstream across the four P:D ratios separates speculation from the transfer path. At light load,
the transfer path alone gives 1.01--1.02$\times$ speedup, so speculation provides most of the gain. The transfer path's
contribution grows with load, mostly from noticing completion immediately (\S\ref{sec:ablation:subsec:streaming_burst}).

Layer-wise writing helps where transfers contend. With speculation disabled and timely completion notification in both
variants, it raises miss-request speedup at 90\% load from 1.103$\times$ to 1.111$\times$ at 1:1 and from 1.094$\times$
to 1.115$\times$ at 3:1, where three prefill instances share each decode instance's link. With interfaces fast enough
that transfers never contend, burst and layer-wise writing give 1.113$\times$ and 1.109$\times$ at 1:1, and
1.097$\times$ and 1.094$\times$ at 3:1. With speculation enabled, burst transfer changes the 1:1 cluster little: miss
requests gain 1.159$\times$ instead of 1.161$\times$ at moderate load and 1.190$\times$ instead of 1.194$\times$ at 90\%
load.

Where P-instances are the bottleneck, speculation adds to the transfer path's gain up to capacity. Where D-instances are the bottleneck, \sys
falls slightly below \sysstream near capacity. Figure~\ref{fig:sim} gives \sysstream the same 0.75\,GiB reserve as \sys,
so that it isolates speculation. At 3:1 capacity, \sys's miss-request speedup over \base is 1.11$\times$, against
1.12$\times$ for \sysstream with the reserve and 1.14$\times$ without it, so the reserve accounts for more than half of
the gap to the unreserved \sysstream.

\subsection{Speedup by Utilization}
\label{app:sim-map}

\begin{figure}[!htb]
\centering
\includegraphics[width=\columnwidth]{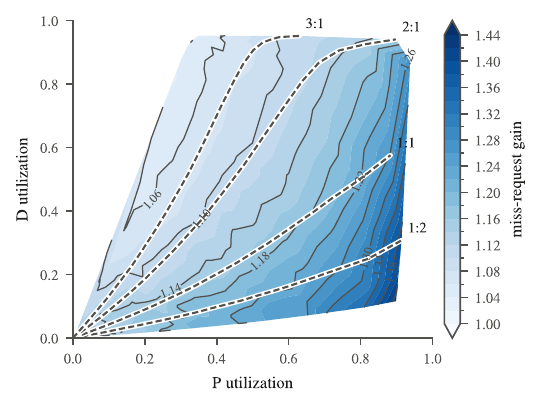}
\caption{Miss-request speedup over P-instance utilization and the D-instance's KV
occupancy (``D utilization'' on the axis), interpolated from
nine P:D ratios. Dashed lines follow the four allocations of
Figure~\ref{fig:sim}.}
\label{fig:sim-map}
\end{figure}

Figure~\ref{fig:sim-map} interpolates nine P:D ratios over P-instance utilization and the D-instance's KV occupancy; its dashed lines follow
the four allocations of Figure~\ref{fig:sim} as their load grows. Miss-request speedups range from 1.04$\times$ to
1.43$\times$, generally increasing with P-instance utilization and decreasing with the D-instance's KV occupancy. The sweeps reach P-instance
utilization of 94\% and D-instance occupancy of 95\%.

\subsection{Deployment Sensitivity}
\label{app:sim-sensitivity}

We vary routing, bandwidth and cluster size around the default configuration (1:1, 96 GPUs, 400\,Gb/s interfaces),
using three arrival schedules per point and each configuration's own measured capacity.

Routing determines whether \sys finds reusable \invkv.
Routers that preserve session affinity, including returning
to the previous instance and cache-aware routing~\cite{sglang},
find the previous request's KV for 93--100\% of miss requests.
Their weekly miss-request speedups are 1.17--1.19$\times$.
Dynamo's KV-cost router~\cite{dynamorouter} finds that KV
for 42\% of miss requests, with another 31\% finding an older
request's KV. It gives 1.12$\times$ weekly speedup and
1.06$\times$ at moderate load. Least-loaded routing finds
the previous KV for 4\% of miss requests, giving 1.06$\times$
weekly speedup and 1.09$\times$ at 90\% load. This routing
also reduces \base's prefix-cache reuse.

Bandwidth and cluster size have smaller effects. Weekly
miss-request speedup ranges from 1.18$\times$ to 1.20$\times$
across NVLink and 100--400\,Gb/s interfaces. It stays at
1.17--1.19$\times$ as the cluster grows from 8 to 192 GPUs
at the same relative load.

\subsection{Combining with Other Techniques}
\label{app:sim-combine}

\sys changes only what the D-instance does while a miss request waits for its exact KV, so it can be combined with techniques that
change where or when that KV is computed. We combine it with two of them in the 96-GPU clusters of \S\ref{sec:sim}:
prefilling short requests on the D-instance, as in Dynamo's conditional disaggregation~\cite{dynamocondagg}, and retaining a session's
KV between its requests~\cite{continuum}. Loads are fractions of \base's
capacity in each allocation, with three arrival schedules per point, and \base alone is the reference.

Dynamo can prefill a request on its decode instance instead of sending it to P: by default, if the request's input beyond
the prefix that the D-instance has cached is under 2{,}048 tokens and under 0.7 of its prompt. Such a request is prefilled on the D-instance in
112-token chunks, never reaches the P-instance and does not speculate. We also try \sys's staging budget as the rule
(Appendix~\ref{app:policy-alternatives}): a request is prefilled on the D-instance if its uncached input $r$ satisfies $rB \le K$, where
$B$ counts the D-instance's running requests and the request. Table~\ref{tab:sim-combine} gives the speedups.

\begin{table}[!htb]
\centering
\caption{\sys combined with prefilling short requests on the D-instance in 96-GPU clusters: all-request speedup over \base, mean of three
arrival schedules. CD: Dynamo's default rule; budget: $rB \le K$. Miss-request speedups differ from these by at most
0.04. $^\dagger$D-instances are overloaded, and not every session finishes.}
\label{tab:sim-combine}
\vskip 0.1in
\footnotesize\setlength{\tabcolsep}{3.1pt}
\begin{tabular}{@{}llrrrr@{}}
\toprule
P:D & Load & \base{}+CD & \sys & \sys{}+CD & \sys{}+budget \\
\midrule
1:2 & 12\% & 1.00 & 1.11 & 1.10 & 1.05 \\
 & 52\% & 1.03 & 1.16 & 1.17 & 1.15 \\
 & 90\% & 1.14 & 1.26 & 1.32 & 1.31 \\
\midrule
1:1 & 12\% & 1.00 & 1.12 & 1.09 & 1.05 \\
 & 52\% & 1.02 & 1.13 & 1.11 & 1.12 \\
 & 90\% & 1.08 & 1.18 & 1.19 & 1.21 \\
\midrule
2:1 & 12\% & 0.99 & 1.11 & 1.08 & 1.04 \\
 & 52\% & 0.97 & 1.08 & 1.03 & 1.08 \\
 & 90\% & 0.76 & 1.12 & 0.83 & 1.10 \\
\midrule
3:1 & 12\% & 0.98 & 1.09 & 1.06 & 1.03 \\
 & 52\% & 0.95 & 1.06 & 0.98 & 1.05 \\
 & 90\% & 0.50$^\dagger$ & 1.09 & 0.52$^\dagger$ & 0.96 \\
\bottomrule
\end{tabular}
\end{table}

Up to 52\% load, Dynamo's rule keeps 28--29\% of miss requests and 80--81\% of the other requests on the D-instance. Without \sys, it
helps only where P-instances are the bottleneck and loaded, and where D-instances are the bottleneck it slows requests down, to 0.50$\times$ at
3:1 and 90\% load; \sys alone is faster at every point. Combined with \sys, the fixed rule adds to the gain where P-instances are
the bottleneck and loaded (1.32$\times$ at 1:2 and 90\% load, against 1.26$\times$ for \sys alone) but costs \sys
elsewhere, since short miss requests prefilled on the D-instance no longer speculate and the D-bound allocations inherit the overload.

\sys's budget as the rule scales with the D-instance's batch, so it keeps more requests on P-instances as D-instances fill. At 90\% load, it gives the
highest speedups of all arms at 1:2 and 1:1, and stays close to \sys alone at 2:1 but not at 3:1 (0.96$\times$ against
1.09$\times$). At light load, it keeps up to 58\% of miss requests on the D-instance, which then do not speculate, and gains only
1.03--1.05$\times$. Prefilling short requests on the D-instance thus complements \sys where P-instances are the bottleneck, provided the rule
accounts for the D-instance's load.

Agent-aware serving systems keep a session's KV cached between its requests. Continuum~\cite{continuum} retains a
finished request's KV for a time-to-live derived from the expected tool time and the cost of recomputing that KV. We
simulate a simpler form with a fixed time-to-live of 30 or 120\,s, during which the D-instance retains a request's prompt and reply and
the P-instance its prompt, unless the session's next request is allocated there first. Retained KV never blocks an allocation, since an
instance without free blocks releases its oldest retentions first. We run the D-bound allocations, 2:1 and 3:1, at 52\%
and 90\% load.

At 52\% load, almost no miss request loses its session's KV (0.3--0.4\% under \sys), and retention changes the speedups by
at most 0.01. At 90\% load, \sys refuses 11.8\% of miss requests because the previous request's KV has been evicted.
Retention lowers this only to 8.7--9.1\%, since retained KV gives way whenever the D-instance needs the memory. The two gains then
compose: at 2:1, retention alone gives 1.02--1.03$\times$, \sys alone 1.12$\times$ and both together 1.14$\times$; at
3:1, 1.01$\times$, 1.09$\times$ and 1.10$\times$.

\end{document}